\documentclass[11pt,a4paper]{article}
\usepackage{jinstpub}

\usepackage[utf8]{inputenc}
\usepackage{graphicx}
\usepackage{placeins}
\usepackage{epstopdf}
\usepackage{todonotes}
\usepackage{lineno}
\usepackage{upgreek}
\usepackage{units}
\usepackage{multirow}
\usepackage{array}
\usepackage{amsmath}
\usepackage{textcomp}
\usepackage{pdfpages}
\usepackage{booktabs}
\usepackage{lscape}
\usepackage{rotating}
\usepackage{bigdelim}
\usepackage{caption}
\usepackage{subcaption}
\usepackage[notlof,notlot,nottoc]{tocbibind}
\usepackage{mathtools}

\newcommand{\cmsq}{cm$^2$}

\title{A Starry Byte -- proton beam measurements of single event upsets and other radiation effects in ABCStar ASIC Versions 0 and 1 for the ITk strip tracker}

\author[a, 1]{M.J.~Basso,}\emailAdd{mbasso@physics.utoronto.ca}
\author[b, c]{J.~Fern\'andez-Tejero,}
\author[d]{B.J.~Gallop,}
\author[b]{G.~Greig,}
\author[e]{J.J.~John,}
\author[f]{P.T.~Keener,}
\author[g]{K.~Krizka,}
\author[e]{P.V.~Leitao,}
\author[h]{B.~Norman,}
\author[d]{P.W.~Phillips,}
\author[b, c, 1]{L.~Poley,}\emailAdd{lpoley@triumf.ca}
\author[d]{C.~Sawyer,}
\author[c]{T.L.~Stack,}
\author[i]{S.~Stucci,}
\author[j]{D.A.~Trischuk,}
\author[k]{and M.~Warren}

\note{Corresponding authors.}

\affiliation[a]{Department of Physics, University of Toronto, Saint George St., Toronto, Canada}
\affiliation[b]{Department of Physics, Simon Fraser University, University Drive W, Burnaby, Canada}
\affiliation[c]{TRIUMF, Wesbrook Mall, Vancouver, Canada}
\affiliation[d]{Particle Physics Department, STFC Rutherford Appleton Laboratory, Harwell Science and Innovation Campus, Didcot, United Kingdom}
\affiliation[e]{Experimental Physics Department, CERN, Geneva, Switzerland}
\affiliation[f]{Department of Physics and Astronomy, University of Pennsylvania, South 33rd Street, Philadelphia, USA}
\affiliation[g]{Lawrence Berkeley National Laboratory, Cyclotron Road, Berkeley, USA}
\affiliation[h]{Department of Physics, Carleton University, Colonel By Drive, Ottawa, Canada}
\affiliation[i]{Brookhaven National Laboratory, Rochester Street, Upton, United States of America}
\affiliation[j]{University of British Columbia, Department of Physics, Agricultural Road, Vancouver, Canada}
\affiliation[k]{Department of Physics and Astronomy, University College London, Gower Street, London, United Kingdom}

\abstract{

Single Event Effects (SEEs) - predominately bit-flips in electronics caused by particle interactions - are a major concern for ASICs operated in high radiation environments such as ABCStar ASICs, which are designed to be used in the future ATLAS ITk strip tracker. The chip design is therefore optimised to protect it from SEEs by implementing triplication techniques such as Triple Modular Redundancy (TMR).

In order to verify the radiation protection mechanisms of the chip design, the cross-section for Single Event Upsets (SEUs), a particular class of SEEs, is measured by exposing the chip to high-intensity particle beams while monitoring it for observed SEUs.

This study presents the setup, the performed measurements, and the results from SEU tests performed using the latest version of the ABCStar ASIC (ABCStar V1) using a \unit[480]{MeV} proton beam.

}


\begin{document}
\maketitle
\flushbottom

\section{Introduction}

For the High-Luminosity Upgrade of the Large Hadron Collider~\cite{HLLHC}, the ATLAS Inner Detector~\cite{ATLAS} will be replaced with the new ATLAS Inner Tracker (ITk) as part of the ATLAS Phase-II Upgrade~\cite{LoI}. The ITk will be an all-silicon tracker and consist of an inner pixel tracker~\cite{TDRp} and an outer strip tracker~\cite{TDRs}.

Modules for the ITk strip tracker consist of silicon strip sensors diced from \unit[6]{inch} wafers, onto which flexes, known as hybrids and powerboards, are glued~\cite{ABC130}. Each sensor strip is read out by a single channel of an ATLAS Binary Chip (ABC) ASIC. ABC ASICs were developed in several iterations (ABCN-25, ABC130, ABCStar~\cite{ABCStar_FE}) to support several generations of module prototyping programme~\cite{ABC130, Petaletpaper}.

The latest version of the ABCStar ASIC, the ABCStar Version~1 (V1), was developed to be used for the construction of the ATLAS detector and designed to be able to operate for an accumulated ionising dose of up to \unit[660]{kGy(Si)} (measured up to \unit[700]{kGy(Si)}). A major consideration for the design revision of the ABCStar Version~0 (V0) was the protection against Single Event Effects (SEEs) by using triplication of registers, clocks, and resets. In order to verify the improved radiation protection mechanisms of the V1 ASIC and its readiness for production, extensive measurements were performed to quantify the cross-sections for Single Event Upsets (SEUs), a class of SEEs corresponding to bit-flips ($0\rightarrow1$ or $1\rightarrow0$), in two proton beam tests. These measurements, which are described in the following sections, comprise the first published results of SEU tests conducted on ABCStar V0 and V1 ASICs.

\section{ABCStar V1}

The ABCStar ASIC is fabricated using \unit[130]{nm} CMOS technology. It is designed to read out signals from 256~sensor strips through binary readout channels. The ABCStar front-end channel is optimised for an input capacitance of \unit[5]{pF}, corresponding to the length of sensor strips on long-strip barrel modules (i.e., \unit[5]{cm}), at a \unit[25]{ns} readout time.

After passing a discriminator, the binary outputs from each front-end channel are sampled and stored in the chip's L0Buffer (effectively a pipeline) for the duration of the programmed latency. The ABCStar architecture supports a multi-trigger data flow:
\begin{itemize}
    \item Level-0 (L0A) trigger \\
    A local version of the Level-0 Accept trigger (and associated ``L0ID'') used in the experiment. Event data is transferred from the beam crossing synchronous pipeline, the L0Buffer.
    \item Low Priority (LP) trigger \\
    A first asynchronous read-out request with priority and low latency for fast processing of track information.
    \item Priority Request (PR) trigger \\
    A second asynchronous read-out request (called PR trigger) intended for a global read-out.
\end{itemize}
When an L0A is received, the event at the end of the L0Buffer pipeline is copied to the EventBuffer memory. The L0A command includes an identification number (called a ``L0tag'', copied from the lowest 7~bits of the L0ID) that is used as the address of the event in the EventBuffer. An LP or PR command will supply a tag that is used to retrieve an event from the EventBuffer memory for processing by the cluster finder. The cluster finder creates clusters of channels where hits were detected, which are then formatted into packets together with the event identification (e.g. L0tag) within the read-out block. A command decoder block receives and distributes trigger signals internally, as well as receiving configuration commands, register read-back instructions, and ASIC settings (e.g. mask and threshold settings). A full description of the ASIC can be found in reference~\cite{TDRs}.

\begin{figure}[htbp]
    \centering
    \includegraphics[width=1.\linewidth]{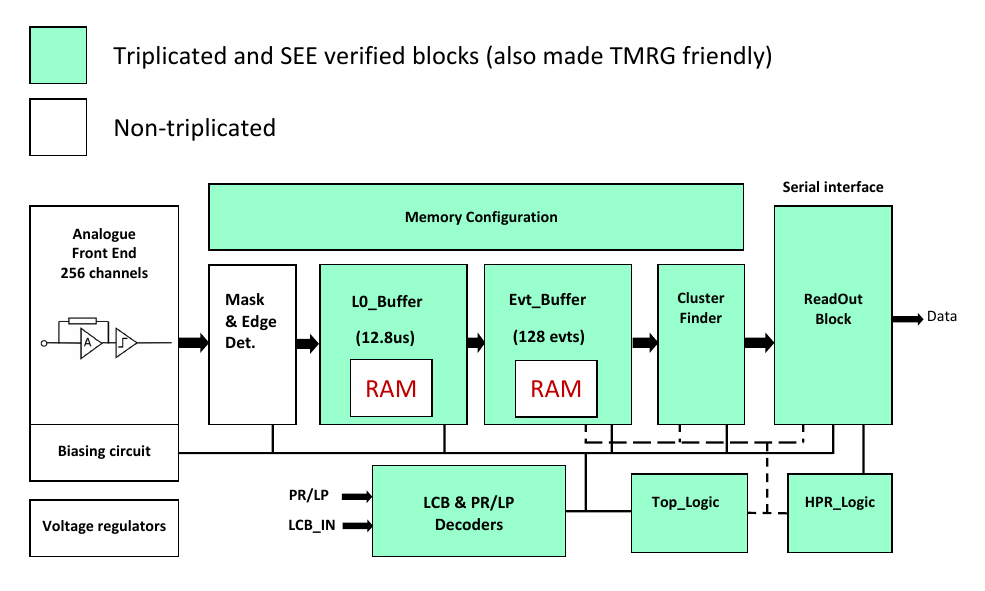}
    \caption{ABCStar V1 block diagram. Please note: ``TMRG'' $\coloneqq$ Triple Modular Redundancy Generator -- for more details, see reference~\cite{TMRG}.}
    \label{fig:triplicated}
\end{figure}

Compared to the ABCStar V0, the SEU protection of the ABCStar V1 is completely revised. Instead of particular registers getting hard-coded protection, the entire project is run through a tool~\cite{TMRG} that generates triplication for all flip-flops and combinatorial logic. Each of the triplicated instances runs from a different clock tree. The remaining parts that are not triplicated are the pipeline memories, the analog multiplexer block that reads ADCs, and the Hit Counter registers (used only for calibration/testing, these count hits per strip allowing occupancy histograms to be recorded on-chip and verification of the readout circuitry). An SEU counter is also included which increments when a corrective action in the triplication logic of specific configuration registers occurs. Figure~\ref{fig:triplicated} shows a block diagram of the ABCStar V1 ASICs indicating what is triplicated and what is not.

Further changes between ABCStar V0 and V1, which are relevant for the studies presented in this paper, include:
\begin{itemize}
    \item SEU counter \\
    An 8-bit SEU counter was added to the ABCStar V1, which increases when an SEU is encountered in any of the following registers: ADCS0, ADCS1, ADCS2, CREG0, or CREG1.\footnote{These registers store information related to ADC monitoring and digital configuration -- details on all of the register types are given in table~\ref{tab:registers}.} The counter does not indicate register errors, but the occurrence of an upset and a subsequent corrective action.
    \item Clock disabling \\
    The ABCStar V1 supports an option to disable one of the three triplicated clocks and thereby the mechanism for internally correcting SEUs occurring in only one of the three instances. Disabling one of the triplicated clocks effectively increases the cross-section by a factor of 2.
    \item Glitch filter \\
    This filter delays a copy of the incoming signal by \unit[1.3]{ns} and compares the copy with the original. If they do not match, the signal is not propagated further. In this way voltage spikes - glitches - of period less than \unit[1.3]{ns} are not seen by the logic beyond this block. The filter is only applied to the configuration pads and to the \unit[40]{MHz} clock input pads on the ABCStar V1 ASIC.
    \item Idle pattern \\
    A programmable idle pattern was introduced to the ASIC, so that the default is not ``0000'' but repeating ``0110'' (like an \unit[80]{MHz} clock).
    \item Programmable multiplexer \\
    For testing purposes, a programmable multiplexer allows an external output pin (``TESTOUT'') to reflect one of a large number of internal signals inside the ASIC.
    \item Packet structure \\
    A comparison of packet structures for the ABCStar V0 and V1 is shown in table~\ref{tab:packets}.

    \begin{sidewaystable}[htbp]
    	\centering
    	\caption{Packet structure for the ABCStar V0 and V1.}
    	\label{tab:packets}
        \resizebox{\textwidth}{!}{
        	\begin{tabular}{|c|c|c|c|c|c|} \toprule
                ABCStar Packet & Physics Packet & V0 Register Read & V1 Register Read & V0 High Priority Register & V1 High Priority Register \\ \midrule
                3 start bits & 3 start bits & 3 start bits & 3 start bits & 3 start bits & 3 start bits \\
                \midrule
                \multirow{4}{*}{Header (16 bits)} & Type: 0001 (PR) / 0010 (LP) & Type:  0100 & Type: 0100 & Type: 1101 & Type: 1101 \\
                                                  & Flag (1 bit) & \multirow{2}{*}{Register address (8 bits)} & \multirow{2}{*}{Register address (8 bits)} & \multirow{2}{*}{Register address (0x3F)} & HPR check bits: 111 \\
                                                  & L0tag (7 bits)            & & & & ``K2 pending'' (1 bit) \\
                                                  & BCID (4 bits)             & 0000 & SEU counter [3:0] & 0000 & SEU counter [3:0] \\
                \midrule
                \multirow{12}{*}{Payload (48 bits)} & \multirow{3}{*}{Cluster 1 (12 bits)} & \multirow{8}{*}{Register contents (32 bits)} & \multirow{8}{*}{Register contents (32 bits)} & \multirow{8}{*}{HPR contents (32 bits)} & \multirow{8}{*}{HPR contents (32 bits)} \\
                 & & & & & \\
                 & & & & & \\
                 & \multirow{3}{*}{Cluster 2 (12 bits)} & & & & \\
                 & & & & & \\
                 & & & & & \\
                 & \multirow{3}{*}{Cluster 3 (12 bits)} & & & & \\
                 & & \multirow{4}{*}{Status bits (16 bits)} & \multirow{4}{*}{Status bits (16 bits)} & \multirow{4}{*}{Status bits (16 bits)} & \multirow{4}{*}{Status bits (16 bits)} \\
                 & & & & & \\
                 & \multirow{3}{*}{Cluster 4 (12 bits)} & & & & \\
                 & & & & & \\
                 & & & & & \\ \midrule
                 1 stop bit & 1 stop bit & 1 stop bit & 1 stop bit & 1 stop bit & 1 stop bit \\ \bottomrule
        	\end{tabular}
        }
    \end{sidewaystable}

    \item Status bits \\
    An overview of the status bits in the ABCStar V0 and V1 is shown in table~\ref{tab:statbits}.

    \begin{table}[htbp]
	\centering
	\caption{Status bits for ABCStar V0 and V1.}
	\label{tab:statbits}
    	\begin{tabular}{r|r|r} \toprule
            V0 status bits & bit & V1 status bits \\ \midrule
            chipID [3] & 15 & chipID [3] \\
            chipID [2] & 14 & chipID [2] \\
            chipID [1] & 13 & chipID [1] \\
            chipID [0] & 12 & chipID [0] \\
            0 & 11 & OR of all SEU bits \\
            BCIDFlag & 10 & Flag (1 bit) \\
            PRFIFO full & 9 & PRFIFO almost full \\
            PRFIFO empty & 8 & PRFIFO empty \\
            LPFIFO full & 7 & LPFIFO almost full \\
            LPFIFO empty & 6 & LPFIFO empty \\
            RegFIFO overflow & 5 & RegFIFO overflow \\
            RegFIFO full & 4 & RegFIFO almost full \\
            RegFIFO empty & 3 & RegFIFO empty \\
            ClusterFIFO overflow & 2 & ClusterFIFO overflow \\
            ClusterFIFO full & 1 & ClusterFIFO almost full \\
            ClusterFIFO empty & 0 & ClusterFIFO empty \\ \bottomrule
    	\end{tabular}
    \end{table}

    \item Test mode \\
    Used during chip testing to apply a set of test patterns to the pipeline in static or pulsed operation modes. In the static mode, the pipeline input continuously gets a set mask bit value.\footnote{N.B. this mode is what allowed for the all 0's and all 1's fills on the front-end of the chip, as described in section~\ref{sec:res_TRIUMF_phys_aug}.} In the pulse mode, there are three controllable options for the input.
\end{itemize}

Table~\ref{tab:registers} shows the different types of the ABCStar registers. The payload of register read packets (32~bits -- see table~\ref{tab:packets}) is the associated register's contents.

\FloatBarrier

\begin{table}[htbp]
	\centering
	\caption{Brief descriptions of the different types of registers present in ABCStar ASICs. Each register is 32 bits deep. When referring to the number of registers of a given type, the bracketed quantity refers to the number present in V0 chips, if different from V1 chips.\protect\footnotemark}
	\label{tab:registers}
    \begin{tabular}{llc} \toprule
        Register type & Description & Number of registers \\ \midrule
        SCReg       & Special/control & 1 \\
        ADCS        & ADC monitoring & 3 (6) \\
        MaskInput   & Mask for each channel & 8 \\
        CREG        & Digital config & 2 (7) \\
        STAT        & Runtime statistics/counters & 7 (5) \\
        HPR         & High priority status codes & 1 \\
        TrimDAC     & Trim for each channel & 40 \\
        CalREG      & Calibration mask & 8 \\
        HitCountREG & Cluster info for each channel & 64 \\ \bottomrule
	\end{tabular}
\end{table}

\footnotetext{N.B. the ABCStar register map was compressed going from V0 to V1, hence the fewer V1 registers.}

\section{Measurement setup}

\begin{figure}[htbp]
    \centering
    \includegraphics[width=0.8\linewidth]{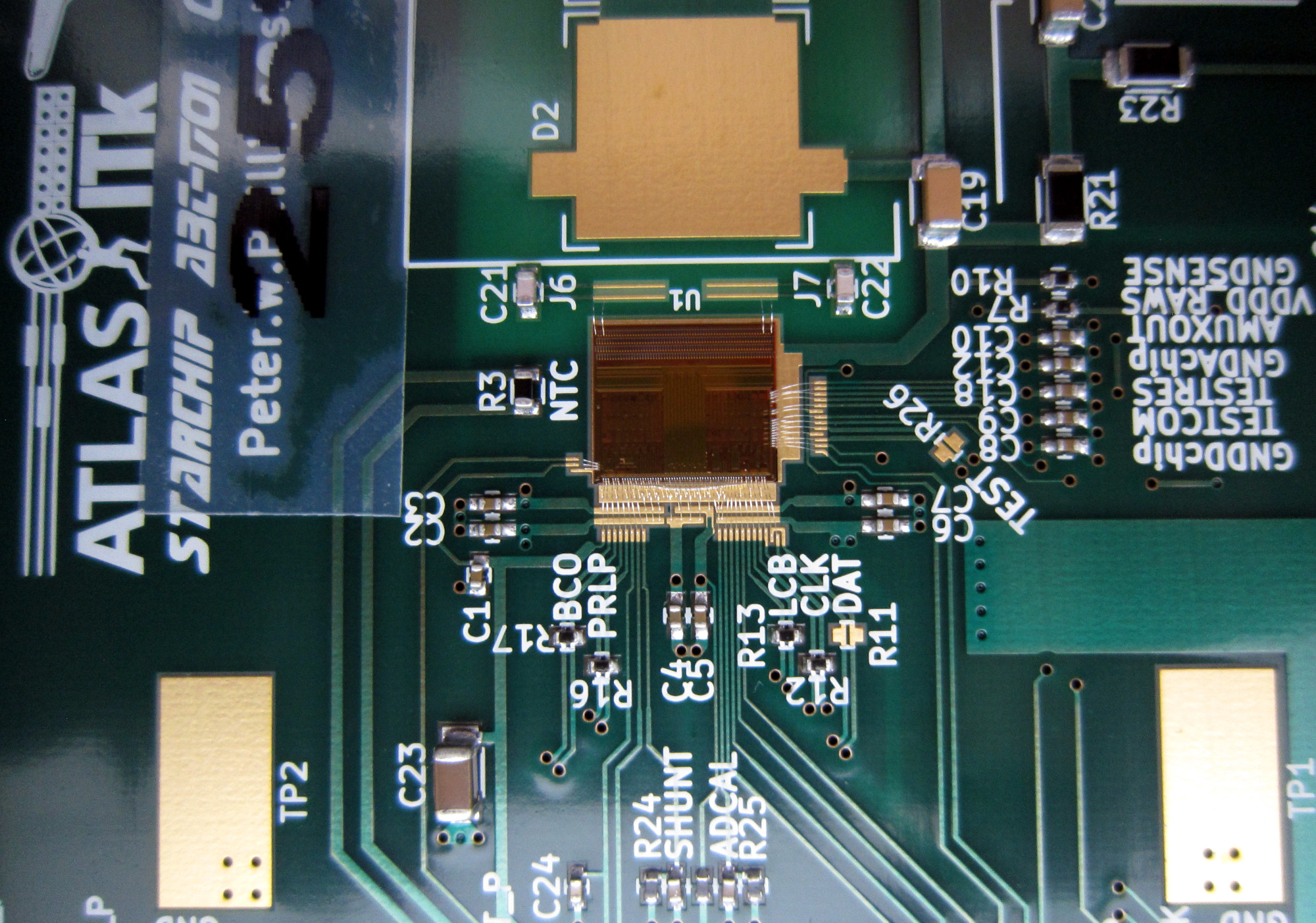}
    \caption{Picture of an ABCStar chip mounted on a Single Chip Board. The ASIC is powered and read out through wire bonds.}
    \label{fig:ABC_SCB}
\end{figure}

For the measurement setup, each ABCStar ASIC was mounted on a dedicated Single Chip Board (SCB) as shown in figure~\ref{fig:ABC_SCB}.

Each SCB is connected to one FMC-1701 printed circuit board (PCB), which is used to power the SCB and facilitate its readout. The cables used to power the SCB were flat ribbon cables, where the current carrying lines had been replaced by thicker wires soldered into the cable. This was done to prevent a voltage drop when the ASIC's current increases due early absorbed dose (an effect called the Total Ionising Dose (TID) bump, see reference~\cite{ABC130}). The chip under investigation was read out using twisted pair cables to minimise cross-talk. The FMC-1701 board was plugged into a Nexys Video FPGA board~\cite{DigilentNV} with an Artix-7 FPGA~\cite{XilinxA7}, which was operated through a PC over a network connection.

During the tests, a constant voltage of \unit[1.5]{V} was applied to each FMC-1701 which, in addition to powering the ASIC, provided a high resolution voltage and current measurement of the supplied voltage and the corresponding digital and analogue current.

\subsection{Setup during measurements}

The Proton Irradiation Facility (PIF)~\cite{PIF} at TRIUMF in Vancouver, Canada provides protons with an energy of \unit[480]{MeV}, which are able to traverse several layers of ASICs on PCBs without significant energy loss. Therefore, multiple boards can be mounted in series and tested. Two beam tests were conducted each using four boards in series. The first, conducted in August 2020, used three ABCStar V1 ASICs and one ABCStar V0 ASIC as a reference.\footnote{Dedicated measurements of the ABCStar V0 SEU rate were conducted during earlier beam tests; however, these results are not publicly available. Here, an ABCStar V0 ASIC was included only to quantify the relative improvement between both ASIC versions.} The second, conducted in December 2020, used only V1 ASICs. Each SCB was connected to one FMC-1701 and one Nexys Video FPGA board, and each FMC-1701 was connected to an individual power supply channel providing \unit[1.5]{V} analogue and digital voltage connected together. All Nexys Video FPGA boards were connected to one Gigabit network switch and read out by one computer.

\begin{figure}[htbp]
    \centering
    \begin{subfigure}{.3\textwidth}
        \centering
        \includegraphics[width=1.\linewidth]{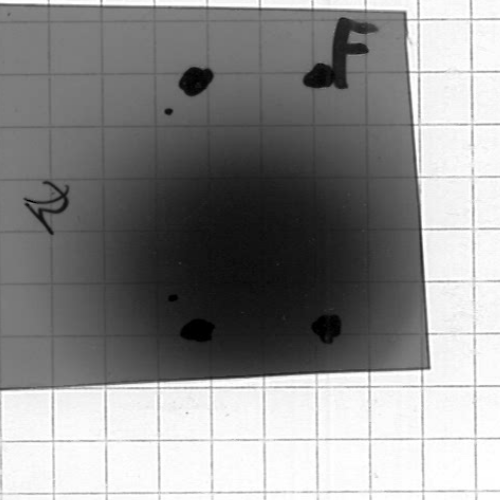}
        \caption{Beam profile at \unit[20]{cm} distance from the end of the beam pipe}
        \label{fig:profile_1}
    \end{subfigure}
    \hfill
    \begin{subfigure}{.3\textwidth}
        \centering
        \includegraphics[width=1.\linewidth]{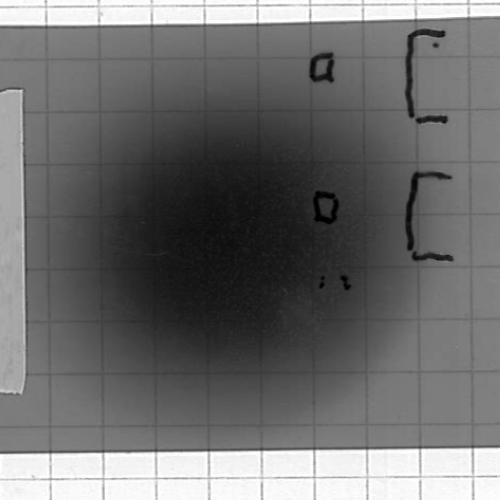}
        \caption{Beam profile at \unit[32]{cm} distance from the end of the beam pipe}
        \label{fig:profile_2}
    \end{subfigure}
    \hfill
    \begin{subfigure}{.3\textwidth}
        \centering
        \includegraphics[width=1.\linewidth]{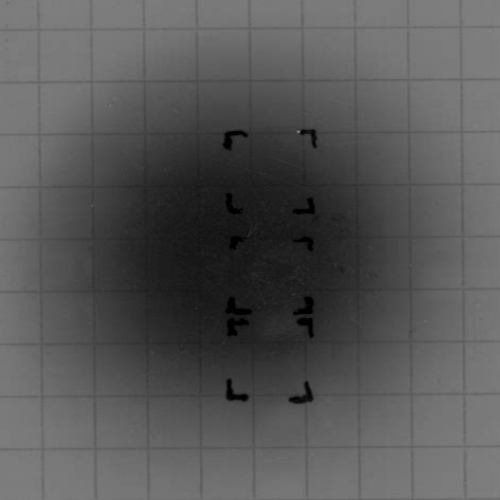}
        \caption{Beam profile at \unit[58.5]{cm} distance from the end of the beam pipe}
        \label{fig:profile_3}
    \end{subfigure}
    \caption{Beam profiles obtained using dosimetry foils, placed at various distances from the end of the beam pipe: the beam diameter increases with increasing distance from the end of the beam pipe. Each square in each grid is \unit[5$\times$5]{mm}.}
    \label{fig:profile}
\end{figure}

Due to an inherent divergence of the proton beam, which increases the beam diameter while decreasing the proton flux per area accordingly, placing the ASICs behind each other exposes them to different dose rates and fluences. Figure~\ref{fig:profile} shows the beam profile measured at different distances from the end of the beam pipe to check the beam diameter.

Dosimetry was performed by measuring the beam flux at different distances to the beam pipe prior to the measurement and extrapolating the flux for each chip based on the particle counter system of the beam line. Tables~\ref{tab:df_Aug} and \ref{tab:df_Dec} summarise the corresponding parameters for all boards mounted in the beam for the August and December beam tests, respectively.

\begin{figure}[htbp]
    \centering
    \includegraphics[width=1.\linewidth]{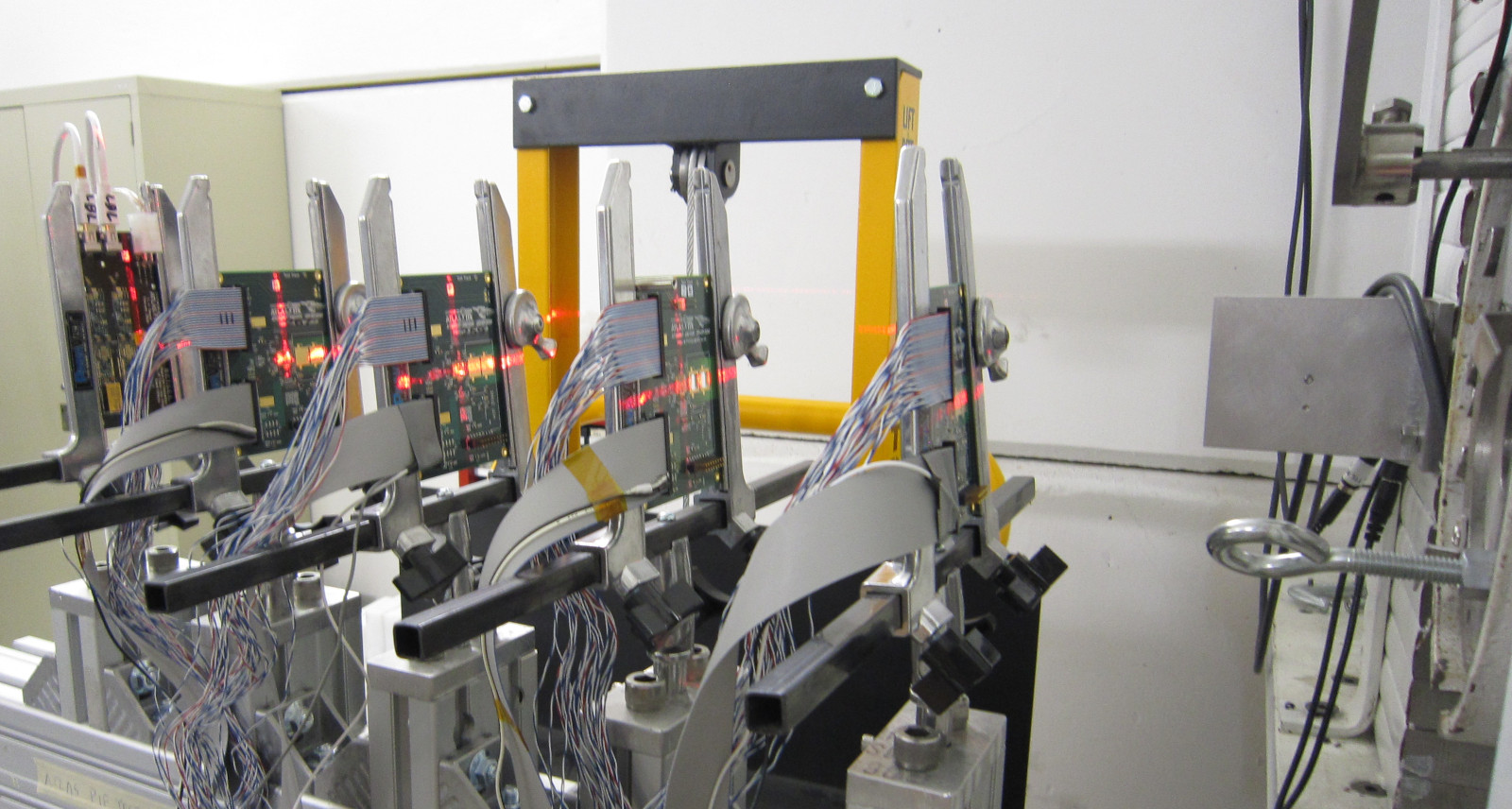}
    \caption{A picture of the setup used in the December 2020 proton beam test. The beampipe is visible on the far right, and the four PCBs to its immediate left correspond to SCBs with ABCStar ASICs mounted. Each SCB is connected to flat and twisted ribbon cables for slow and fast signals, respectively. The red laser focused on the ABCStar ASICs indicates the position of the beam.}
    \label{fig:proton_setup}
\end{figure}

A picture of the setup used in the August beam test is shown in figure~\ref{fig:proton_setup}. A similar setup was used for the December beam test.

\begin{table}[htbp]
	\centering
	\caption{Doses and fluences accumulated by all boards under investigation as part of the August 2020 proton beam test. The parameters of each board are given as a function of their distance to the beam pipe.}
	\label{tab:df_Aug}
	\begin{tabular}{ccrrr} \toprule
        Chip & Generation & Distance & Dose & Fluence \\
        & & [cm] & [Mrad] & [p/\cmsq] \\ \midrule
        261 & V1 & 18.8 & 4.30 & $1.18\times10^{14}$ \\
        002 & V0 & 31.0 & 3.34 & $9.17\times10^{13}$ \\
        259 & V1 & 44.0 & 2.67 & $7.33\times10^{13}$ \\
        267 & V1 & 56.3 & 2.23 & $6.13\times10^{13}$ \\ \bottomrule
	\end{tabular}
\end{table}

\begin{table}[htbp]
	\centering
	\caption{Doses and fluences accumulated by all boards under investigation as part of the December 2020 proton beam test. The parameters of each board are given as a function of their distance to the beam pipe.}
	\label{tab:df_Dec}
	\begin{tabular}{ccrrr} \toprule
        Chip & Generation & Distance & Dose & Fluence \\
        & & [cm] & [Mrad] & [p/\cmsq] \\ \midrule
        259 & V1 & 43.5 & 2.48 & $6.80\times10^{13}$ \\
        273 & V1 & 46.0 & 2.38 & $6.55\times10^{13}$ \\
        267 & V1 & 56.0 & 2.07 & $5.67\times10^{13}$ \\
        274 & V1 & 58.5 & 2.00 & $5.49\times10^{13}$ \\ \bottomrule
	\end{tabular}
\end{table}

\FloatBarrier

\section{Measurements}

During the beam tests, ABCStars were read out using the Inner Tracker Strips DAQ (ITSDAQ) software~\cite{ABC130}. The Nexys Video FPGA board was loaded with ITSDAQ firmware (FW)~\cite{ITSDAQfw} version \texttt{vb48c} for the August beam test and firmware version \texttt{vb4b4} for the December beam test.\footnote{The important differences going from firmware version \texttt{vb48c} to \texttt{vb4b4} include support for idle mode, the addition of TESTOUT counters, and the expansion of the timestamp register rollover size -- it was increased from 32~bits to 48~bits (i.e., \unit[107]{s} to \unit[$\sim\!2000$]{hours}).}

In order to determine the SEU rates, the following data was monitored:
\begin{itemize}
    \item register data (triplicated, for writable as well as read-only registers);
    \item physics data (non-triplicated event data);
    \item TESTOUT output counters.
\end{itemize}

The firmware TESTOUT counters are connected to the ABCStar TESTOUT debug output that can be switched to monitor a selection of signals inside the ASIC. Some of these may be affected by SEUs. To monitor this activity, three counters are available in the firmware. The first counts the number of times the TESTOUT signal changes from 0 to 1, the second counts the integrated period that the signal is asserted, and third is continuously running and can be used as a time reference. Firmware status information is read out as a block 16-bit words, with the TESTOUT counters wider than 16-bit assembled from multiple of these words:
\begin{itemize}
    \item Edge Counter: 32-bit.

    Counts rising-edges in TESTOUT at \unit[1.56]{ns} (\unit[1/640]{MHz}) resolution.\footnote{The underlying logic operates on bytes at \unit[80]{MHz}, so the maximum number of rising edges possible in a byte is 4 (N.B. the last bit of the previous byte is also used to check for an edge between the two). The max value is 0xfffffffb+last byte edge-count. This means the max value is the range 0xfffffffc-f.}

    \item Period-high Counter: 32-bit.

    Counts the period that TESTOUT is high in \unit[1.56]{ns} steps.\footnote{
    Similar to Edge Count, the maximum number of steps a byte can be high is 8, the max value is 0xfffffff7+last byte count. This means the max value is the range 0xfffffff8-f.}

    \item Timecount Counter: 48-bit.

    Free running counter, clocked by a \unit[40]{MHz} clock, set to 0 at power-up/reset.

\end{itemize}

Only the ``LCB\_Locked'' signal was monitored. The LCB (\textbf{L}0A \textbf{C}ommand \textbf{B}unch Counter Reset) input to the ABCStar carries L0A\footnote{L0A: Level 0 Accept: beam-synchronous pulse that transfers data from L0Buffer to the EventBuffer.}, Command\footnote{Command: controls register reading and writing.} and BCR\footnote{BCR: Bunch Counter Reset: pulse synchronous with the machine orbit used to keep local counter aligned} signals. These are used to control the ASIC, including register reading and writing. The LCB is transferred in 16-bit frames at \unit[160]{Mb/s}, to which the logic in the ABCStar must synchronise (or ``lock'') to. Once a sufficient number of good frames are seen, the LCB decoder enters a ``locked'' state. The LCB signal is always active and encoded such that errors can easily be identified, and persistent errors will cause the ABCStar logic to de-assert ``locked''. As such, this is either a good indicator of a serious failure in this part of the ASIC or that a reset has occurred.

Monitoring was grouped into several loops:
\begin{enumerate}
    \item An inner loop filling and reading out the event data buffer;
    \item A middle loop, during which:
    \begin{itemize}
        \item the inner loop was executed, in sets of 10, in the case of the August beam test, or 100, in the case of the December beam test;
        \item the register data was read out and writable registers were reset (the code was optimised to maximise integration time by resetting the register directly after reading out);
        \item the monitoring data from the TESTOUT counters and the FMC-1701 was read out and stored.
    \end{itemize}
    Each step was logged with a timestamp for subsequent calculation of integration time and corresponding fluence.
    \item An outer loop: in the case of the August beam test, the fill pattern was changed for sets of 10 middle loops (``fixed'' versus BCID-dependent fills -- see section~\ref{sec:res_TRIUMF_phys_aug} for more details).
\end{enumerate}

Additionally, several options and running modes were implemented:
\begin{itemize}
    \item Triplication enabled $|$ disabled \\
    The occurrence of SEUs in registers is rare due to their triplication. In order to verify that the setup was able to detect SEUs, a test mode was implemented where the third clock was disabled and the rate of observable SEUs increased. This mode was used for initial tests.
    \item Glitch filter mode enabled $|$ disabled \\
    A glitch filter was added to the configuration pads and to the \unit[40]{MHz} clock input pads on the ABCStar V1 ASIC. The goal is to filter out glitches that could come from external sources to the chip. The glitch filter was activated for half of the data collection time in order to compare the SEU cross-sections with and without glitch filter enabled.
    \item Fill pattern fixed $|$ BCID \\
    Two modes of filling the ABCStar's event data were tested: ``Fixed'', where all hits were filled with 1's or 0's (alternating per outer loop), and ``BCID'', where, during the inner loop, the L0Buffer was filled with all possible combinations of 4-strip clusters in the event buffer. While the ``Fixed'' pattern allowed for straightforward of comparison of the cross-sections for 0$\rightarrow$1 and 1$\rightarrow$0 bit flip cross-sections and straightforward analysis, the BCID pattern data provided a more realistic approximation of real event data.
    \item Latency setting \\
    The ABCStar is a pipelined device, where data is stored for some period of time awaiting a trigger. This setting allows the time in which data stays in the pipeline (in clock periods) to be changed, which means more SEUs might be integrated in this time.
\end{itemize}

A second mode of data collection was implemented for the December 2020 proton measurements called ``Idle'' mode: instead of gathering physics data, the ASIC was configured to send an ``Idle'' pattern - that is a repetitive pattern on the data output - of ``0110''. In the absence of SEUs, data from the ASIC was expected to contain only the idle pattern and, at scheduled intervals, high-priority packets (HPRs). If unexpected packets were seen, as would happen if, for example, the ASIC saw a trigger, this would be known to be caused by SEEs.

\section{Results from proton tests}
\label{sec:res_TRIUMF}

The following sections describe the measurements performed as part of the proton beam tests at TRIUMF in August and December 2020. Where appropriate, the integrated fluence or dose for a given chip is calculated using the calibration curves and particle counts provided by the TRIUMF PIF, following the procedure described in appendix~\ref{app:res_TRIUMF_dose_meas}.

\subsection{Firmware TESTOUT counters}
\label{sec:res_TRIUMF_testout}

\begin{figure}[htbp]
    \centering
    \begin{subfigure}{.49\textwidth}
        \centering
        \includegraphics[width=1.\linewidth]{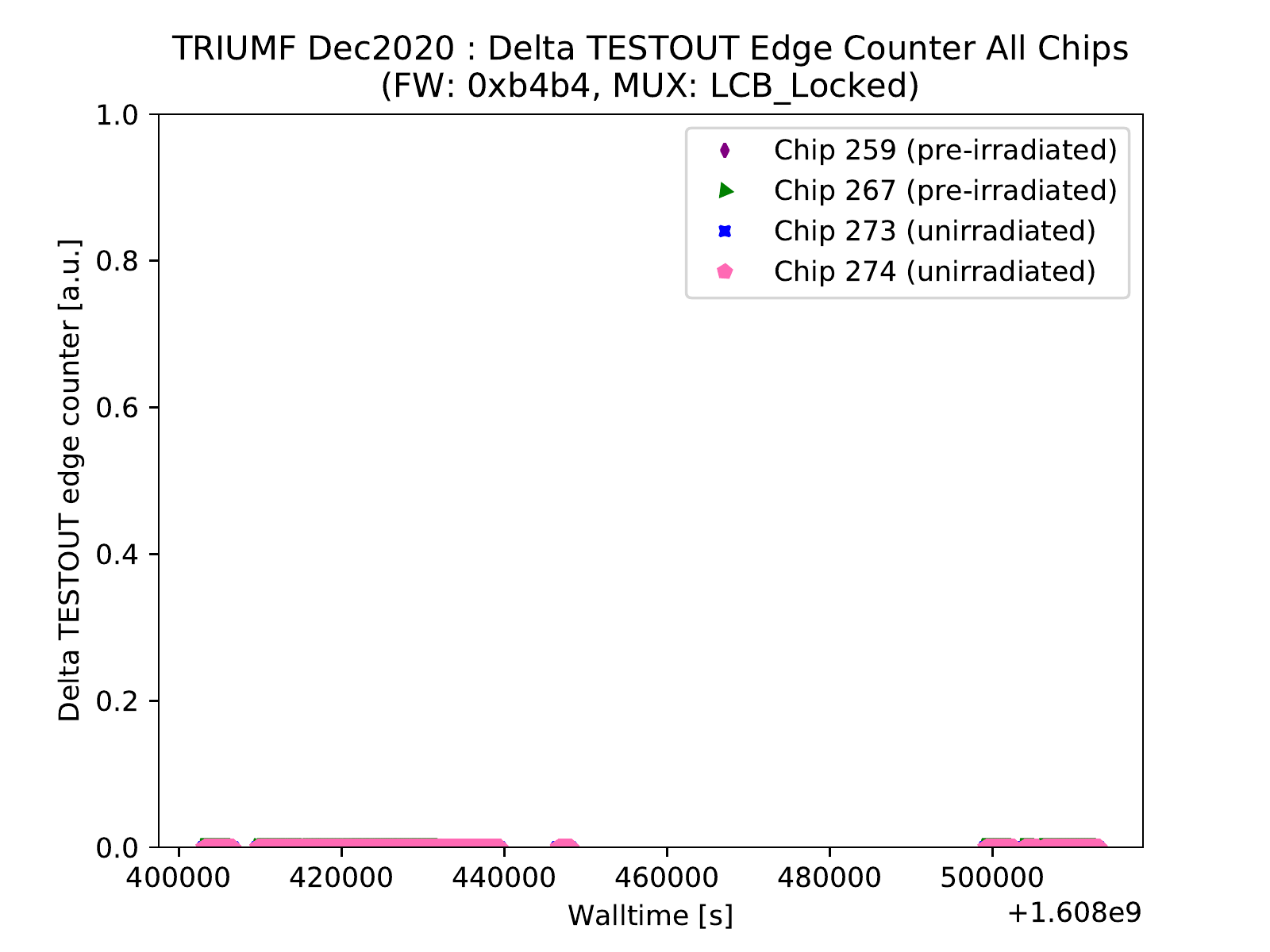}
        \caption{Edge counter}
        \label{fig:proton_testout_edge}
    \end{subfigure}
    \hfill
    \begin{subfigure}{.49\textwidth}
        \centering
        \includegraphics[width=1.\linewidth]{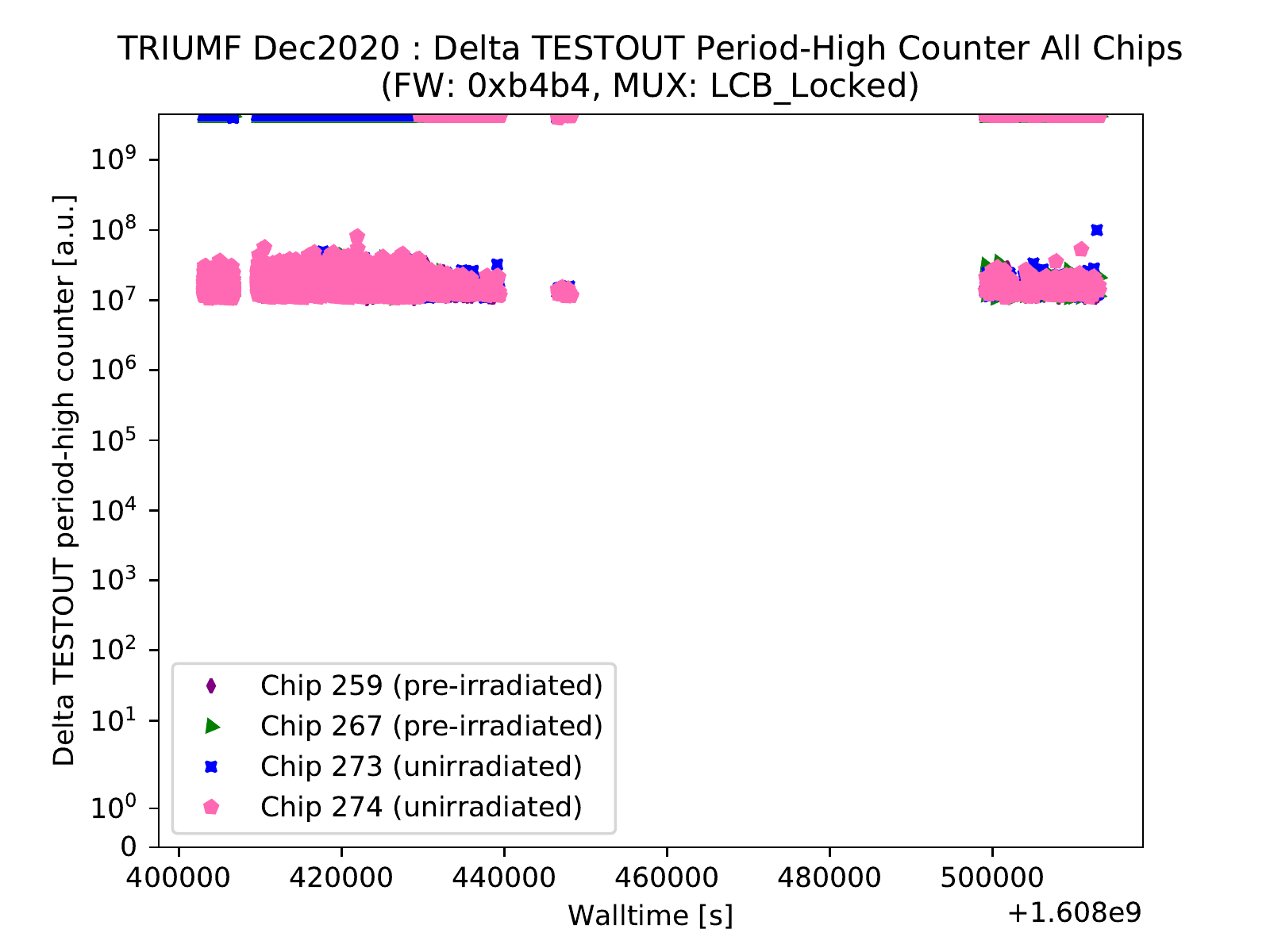}
        \caption{Period-high counter}
        \label{fig:proton_testout_period_high}
    \end{subfigure} \\
    \centering
    \begin{subfigure}{.49\textwidth}
        \centering
        \includegraphics[width=1.\linewidth]{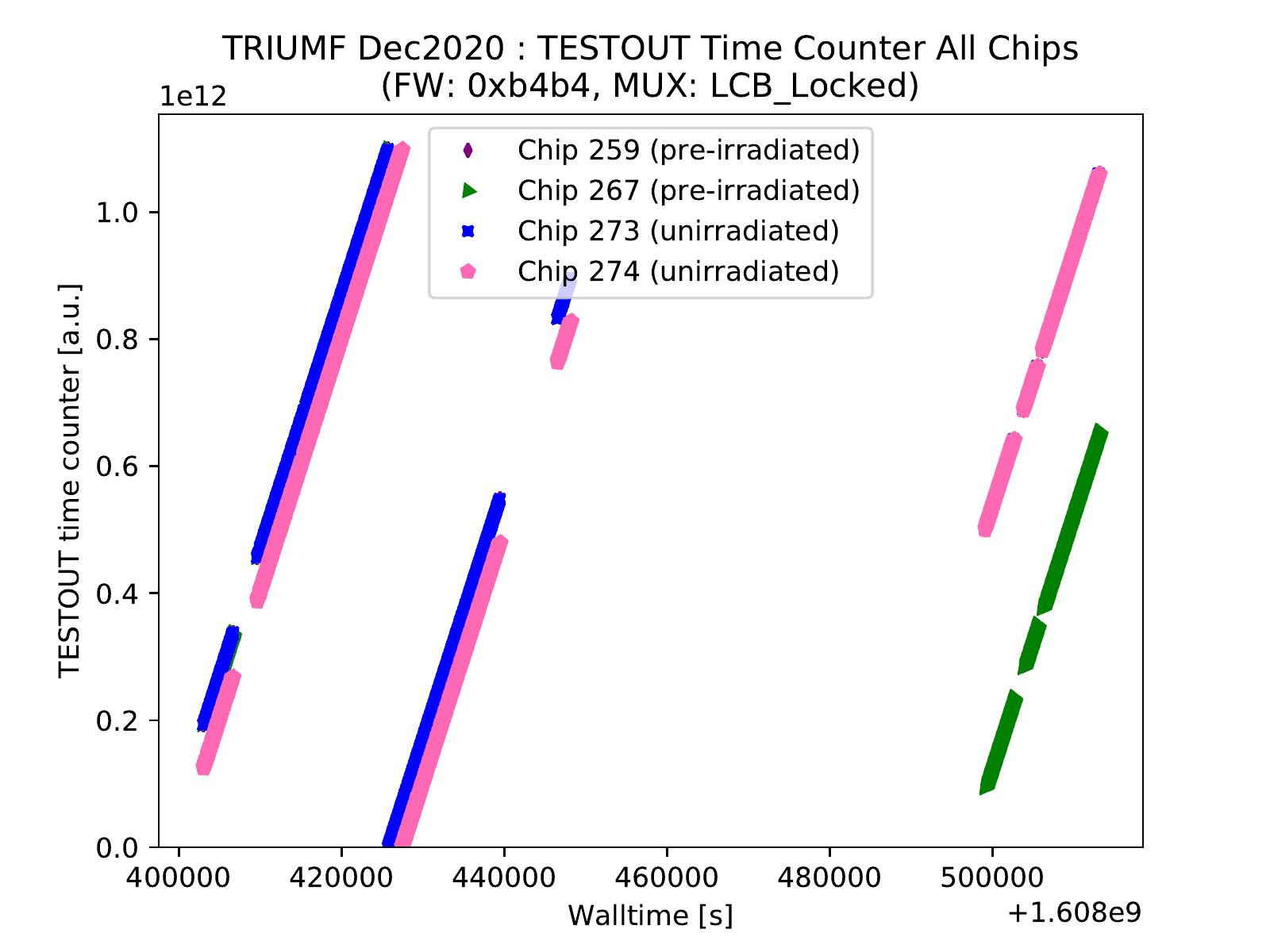}
        \caption{Time counter}
        \label{fig:proton_testout_timecount}
    \end{subfigure} \\
    \caption{Firmware TESTOUT counters for the ABCStar V1 ASICs measured as part of the December 2020 beam test. ``0xb4b4'' refers to the firmware revision (\texttt{vb4b4}) used. N.B. the large time gap between some data points corresponds to the time between days 1 and 2 of the beam test. Additionally, the vertical scale for the period-high counter, (b), is logarithmic with a linear scale near 0.}
    \label{fig:proton_testout}
\end{figure}

The results from the ABCStar’s TESTOUT counters for the December beam test are shown in figure~\ref{fig:proton_testout}. Only the ``LCB\_Locked'' signal was monitored. N.B. the TESTOUT counter functionality was \emph{not} included as part of the August beam test.

From figure~\ref{fig:proton_testout_edge}, no edges were counted -- this indicates the LCB\_Locked signal output never transitioned high-to-low or low-to-high. And from figure~\ref{fig:proton_testout_period_high}, the LCB\_Locked signal output is always high, as the period-high counter is always at or near saturation following a counter reset. Altogether, these results indicate that \emph{no} LCB unlocked events were measured for any of the ASICs over the duration of the beam test, indicating no serious failures.

From figure~\ref{fig:proton_testout_timecount}, the time counted in firmware is proportional to the time counted in software and the slope of the linear relationship between firmware and software time is the same for all chips, as expected.

\subsection{HPR reset and unlock rates}
\label{sec:res_TRIUMF_hpr}

High Priority Register (HPR) packets are normal packets containing link status information that are sent with high priority at regular intervals. They are initiated with a higher priority than any other packet type. There are two rate measurements of interest made with respect to HPR packets: the rate of HPR resets and the rate of LCB unlocks.

During normal operation, the HPR packets are expected to be spaced apart in integer multiples of 40,000~bunch crossings (BCs), corresponding to integer multiples of \unit[1]{ms}. If an HPR packet occurs at another time after the previous HPR packet, then this would be indicative of a reset in the ASIC. However, if an HPR is scheduled to occur while another packet is already being read out (or a group of packets), the time when the HPR packet is read out may be modified by the time required to read out the conflicting packet (or packets). Therefore, the check for out-of-time packets is $|\Delta\textrm{BC}\!\mod 40,000| > 67$~BCs, where 17 out of 67~BCs are for packet readout\footnote{There are 68~bits per packet and bits are read at a rate of 4~bits per BC, yielding 17~BCs to read the entire packet.} and 50 out of 67~BCs are to account for further delays due to the readout of additional physics packets initiated by SEUs. For events in August with all 1's L0Buffer fills, if an HPR occurs during the LP or PR buffer scans, 17~BCs is modified to $16\times17=272$~BCs as there are 16~packets read from the LP or PR buffers per scan per L0A (as opposed to 1~packet for events with all 0's fills).

One of the status bits contained in the HPR is the ``LCB\_Locked'' (bit 29). The number of LCB unlocks can therefore be determined by counting the number of times this bit is seen to switch from ``1'' to ``0''. In this section as well as section~\ref{sec:res_TRIUMF_status}, pattern matching for ``good'' HPR status bits (``0101001001'' -- these bits correspond to the ``PRFIFO full/almost full'' through ``ClusterFIFO empty'', inclusive, status bits in table~\ref{tab:statbits} followed by a stop bit followed by a ``0000'' or ``0110'' idle sequence) is used to guard against unlocked HPR packets which are actually mis-parsed, locked HPR packets.

For both the numbers of out-of-time and unlocked HPR packets, the rates are calculated by summing the integrated times between subsequent packets.

\subsubsection{August results}
\label{sec:res_TRIUMF_hpr_aug}

The HPR packet results for the August 2020 beam test are summarised in table~\ref{tab:proton_hpr_aug}. With the exception of the BCID fill runs with the glitch filter enabled for Chips 261 and 267 (which see $\mathcal{O}(0.5)$~packets/s), all of the V1 chips see $\mathcal{O}(5)$~packets/s for all runs. In comparison, the V0 chip sees $\mathcal{O}(1)$~packets/s for all runs.

\emph{No} unlocked packets were measured for any runs. A 95\% upper confidence bound is calculated by assuming 3 unlocked packets in the rate calculation -- these upper bounds are $1.67 \times 10^{-5}$~packets/s and $5.04 \times 10^{-6}$~packets/s for the V0 and V1 chips, respectively.

\begin{figure}[htbp]
    \centering
    \includegraphics[width=0.8\linewidth]{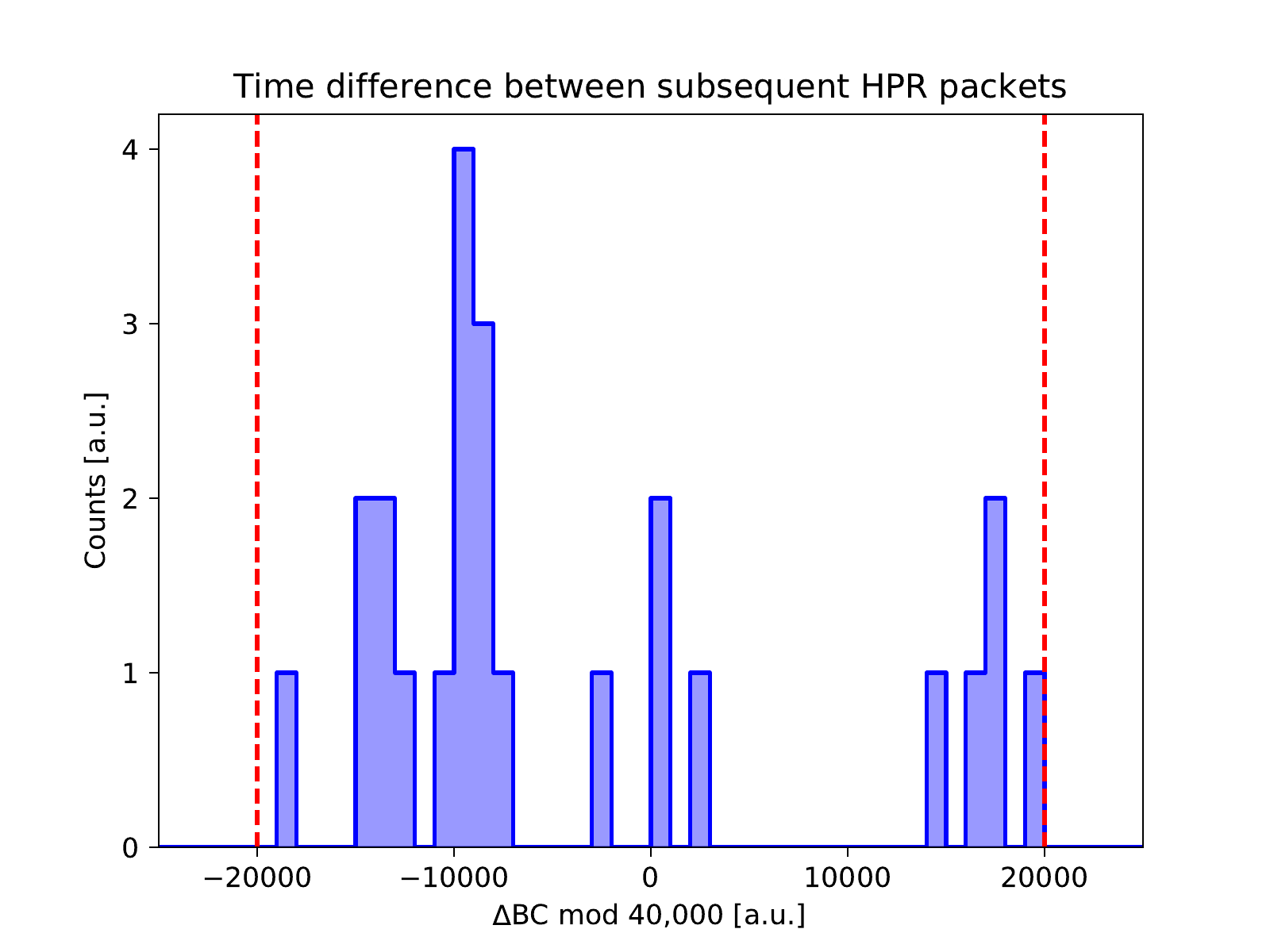}
    \caption{Histogram of delta times (in BCs) between subsequent HPR packets for the out-of-time packets measured in August 2020. The dotted red lines denote the domain of the modulo~40,000 operation (i.e., $\pm$20,000~BCs).}
    \label{fig:hpr_delta_times}
\end{figure}

Altogether, 17 and 7~$\Delta{t} \neq 1$~ms packets were seen for the V0 and V1 chips, respectively. In general, $\Delta\textrm{BC} \!\mod 40,000$ is large for the out-of-time packets -- the distribution of $\Delta\textrm{BC} \!\mod 40,000$ for out-of-time packets for all chips is shown in figure~\ref{fig:hpr_delta_times}. As none of the chips showed a corresponding LCB unlock event or a register with unexpected contents, these are assumed to be related to the data-taking mode (with an extra reset as part of setup) rather than an unscheduled ASIC reset. Therefore, this extra reset was removed in the following run. We may still calculate the rate of $\Delta{t} \neq 1$~ms packets as:

\begin{equation}
    \textrm{Rate}_{\Delta{t} \neq 1\,\textrm{ms}} = \frac{N_{\textrm{packets},\Delta{t} \neq 1\,\textrm{ms}}}{\int dt} \,,
\end{equation}

\noindent and the 95\% confidence interval on the rate as:

\begin{equation}
    \Delta_{\textrm{Rate}_{\Delta{t} \neq 1\,\textrm{ms}}} = 1.96 \times \frac{\sqrt{N_{\textrm{packets},\Delta{t} \neq 1\,\textrm{ms}}}}{\int dt} \,,
\end{equation}

\noindent (assuming Poisson counting statistics). The factor of 1.96 comes from $\Phi(1 - \alpha/2) = \Phi(1 - 0.05/2) \approx 1.96$ where $\Phi$ is the quantile function of a standard normal distribution. The rates are calculated to be $(9 \pm 4) \times 10^{-5}$~packets/s and $(1.7 \pm 1.2) \times 10^{-5}$~packets/s for the V0 and V1 chips, respectively.

\begin{table}[htbp]
    \centering
	\caption{HPR packets results for August 2020. Both the number of unlocked HPR packets ($N_\textrm{unlocked}$) and the number HPR packets with delta times not equal to 1 ms ($N_{\Delta{t} \neq 1\,\textrm{ms}}$) are shown.}
	\label{tab:proton_hpr_aug}
    \resizebox{\textwidth}{!}{
    	\begin{tabular}{clllrrrr} \toprule
            Chip & Fill type & Glitch filter? & Clock disabled? & $N_\textrm{unlocked}$ & $N_{\Delta{t} \neq 1\,\textrm{ms}}$ & $N_\textrm{packets}$ & Integrated time [s] \\\midrule
            002 & Fixed & No  & No  & 0 & 17 & 171,557 & 179677.0 \\\midrule
            \multirow{5}{*}{259} & BCID  & No  & No  & 0 &  1 &  65,787 &  15263.7 \\
            & BCID  & Yes & No  & 0 &  0 &  66,211 &  15665.1 \\
            & Fixed & No  & No  & 0 &  0 &  96,858 &  18142.5 \\
            & Fixed & No  & Yes & 0 &  0 &   3,871 &    801.7 \\
            & Fixed & Yes & No  & 0 &  0 &  84,721 &  15930.2 \\\midrule
            \multirow{5}{*}{261} & BCID  & No  & No  & 0 &  1 &  68,390 &  15825.9 \\
            & BCID  & Yes & No  & 0 &  1 &  73,978 & 127189.0 \\
            & Fixed & No  & No  & 0 &  0 &  82,822 &  16263.3 \\
            & Fixed & No  & Yes & 0 &  0 &   7,102 &   1386.7 \\
            & Fixed & Yes & No  & 0 &  0 &  86,364 &  15949.9 \\\midrule
            \multirow{5}{*}{267} & BCID  & No  & No  & 0 &  0 &  74,046 &  17007.5 \\
            & BCID  & Yes & No  & 0 &  1 &  28,601 & 117843.0 \\
            & Fixed & No  & No  & 0 &  0 &  67,774 &  13509.9 \\
            & Fixed & No  & Yes & 0 &  1 &   7,283 &   1390.6 \\
            & Fixed & Yes & No  & 0 &  2 & 132,798 &  23811.3 \\
            \bottomrule
    	\end{tabular}
    }
\end{table}

\subsubsection{December results}
\label{sec:res_TRIUMF_hpr_dec}

The HPR packet results for the December 2020 beam test are summarised in table~\ref{tab:proton_hpr_dec}. \emph{No} unlocked packets or packets with delta times out of specification were measured -- this is consistent with the FW counter results described in section~\ref{sec:res_TRIUMF_testout}: no ASIC resets were measured.

We can calculate 95\% upper confidence bounds on the rate of LCB unlocked and $\Delta{t} \neq 1\,\textrm{ms}$ (altogether: ``bad'') packets. Assuming the results are not dependent on the chip's latency, upper bounds on the rate of bad HPR packets for the normal and idle running modes are $1.94 \times 10^{-5}$~packets/s and $1.29 \times 10^{-4}$~packets/s, respectively. Here, we have distinguished between the normal and idle running modes as the normal mode sees $\mathcal{O}(100)$~packets/s while the idle mode sees $\mathcal{O}(500)$~packets/s.

While the ASIC specifications do not place an explicit limit on these effects, the observed rates for both August and December were found to be of no concern for the operation of ABCStar ASICs in the detector.

\begin{table}[htbp]
    \centering
	\caption{HPR packets results for December 2020. Both the number of unlocked HPR packets ($N_\textrm{unlocked}$) and the number HPR packets with delta times not equal to 1 ms ($N_{\Delta{t} \neq 1\,\textrm{ms}}$) are shown.}
	\label{tab:proton_hpr_dec}
    \resizebox{\textwidth}{!}{
    	\begin{tabular}{clrrrrr} \toprule
            Chip & Idle mode? & Latency & $N_\textrm{unlocked}$ & $N_{\Delta{t} \neq 1\,\textrm{ms}}$ & $N_\textrm{packets}$ & Integrated time [s] \\\midrule
            \multirow{2}{*}{259} & No  &  20 & 0 & 0 &  3,323,805 & 33173.1 \\
            & No  & 503 & 0 & 0 &  1,395,467 & 12312.2 \\\midrule
            \multirow{2}{*}{267} & No  &  20 & 0 & 0 &  2,591,380 & 25863.8 \\
            & No  & 503 & 0 & 0 &  1,417,932 & 12313.0 \\\midrule
            \multirow{2}{*}{273} & No  &  20 & 0 & 0 &  3,543,084 & 34829.8 \\
            & No  & 503 & 0 & 0 &  1,418,156 & 12313.7 \\\midrule
            \multirow{3}{*}{274} & Yes &  20 & 0 & 0 & 12,101,755 & 23227.2 \\
            & No  &  20 & 0 & 0 &  1,015,722 &  9888.1 \\
            & No  & 503 & 0 & 0 &  1,617,822 & 13964.5 \\
            \bottomrule
    	\end{tabular}
    }
\end{table}

\subsection{HPR and register read status bits}
\label{sec:res_TRIUMF_status}

The status bits for HPR and register read packets are described in table~\ref{tab:statbits}. For all HPR and register read packets in August and December, the status bits were checked for FIFO almost-full (``*FIFO\_almost\_full'') or, in the case of V0 chips, full statuses (``*FIFO\_full'') as well as for register or cluster FIFO overflows, which could indicate a reset in the ASIC.

\subsubsection{August results}
\label{sec:res_TRIUMF_status_aug}

The status bit results for the August beam test are summarised in table~\ref{tab:proton_stat_aug}. Chip 002, the V0, returned 159 and 228 PRFIFO\_full and LPFIFO\_full statuses, respectively, corresponding to 0.0007\% and 0.001\% of all packets measured. Chip 261, a V1, returned 18 LPFIFO\_almost\_full statuses for its BCID fill run, corresponding to 0.00007\% of all packets measured. Chip 267, a V1, returned 150 ClusterFIFO\_almost\_full statuses for its clock-disabled run, corresponding to 0.0005\% of all packets measured. Additionally, \emph{no} FIFO overflow statuses were measured -- this is consistent with the HPR out-of-time results described in section~\ref{sec:res_TRIUMF_hpr_aug}: no ASIC resets were measured.

An attentive reader may notice that the number of HPR and register read packets measured for the fixed L0Buffer fill runs do not sum to the number of expected packets given table~\ref{tab:proton_hpr_aug} and (later, in section~\ref{sec:res_TRIUMF_reg_aug}) table~\ref{tab:proton_reg_xs_aug}. This is because the fixed L0Buffer fill runs perform full LP/PR buffer scans, leading to a full readout of the HitCountREG registers (\texttt{0x80} through \texttt{0xbf}, inclusive). As a result, an additional 64~register read packets per buffer read per scan type are measured for these runs.

\begin{sidewaystable}[htbp]
    \centering
	\caption{HPR and register read status bits for the August 2020 beam test. The status bit columns indicate the number of packets which had 1 instead of 0 for that status bit. For Chip 002, the V0, the ``almost\_full'' FIFO statuses actually correspond to ``full'' statuses. N.B. a packet may count towards the yields in more than one status bit column.}
	\label{tab:proton_stat_aug}
    \resizebox{\textwidth}{!}{
    	\begin{tabular}{clllccccccr} \toprule
            Chip & Fill type & Glitch filter? & Clock disabled? & PRFIFO\_almost\_full & LPFIFO\_almost\_full & RegFIFO\_overflow & RegFIFO\_almost\_full & ClusterFIFO\_overflow & ClusterFIFO\_almost\_full & $N_\textrm{packets}$ \\\midrule
            002 & Fixed & No  & No  & 159 & 228 & 0 & 0 & 0 &   0 & 22,916,647 \\\midrule
            \multirow{5}{*}{259} & BCID  & No  & No  &   0 &   0 & 0 & 0 & 0 &   0 &    179,107 \\
            & BCID  & Yes & No  &   0 &   0 & 0 & 0 & 0 &   0 &    180,236 \\
            & Fixed & No  & No  &   0 &   0 & 0 & 0 & 0 &   0 & 14,624,436 \\
            & Fixed & No  & Yes &   0 &   0 & 0 & 0 & 0 &   0 &    596,129 \\
            & Fixed & Yes & No  &   0 &   0 & 0 & 0 & 0 &   0 & 12,374,915 \\\midrule
            \multirow{5}{*}{261} & BCID  & No  & No  &   0 &  18 & 0 & 0 & 0 &   0 &    186,872 \\
            & BCID  & Yes & No  &   0 &   0 & 0 & 0 & 0 &   0 &    202,725 \\
            & Fixed & No  & No  &   0 &   0 & 0 & 0 & 0 &   0 & 12,438,698 \\
            & Fixed & No  & Yes &   0 &   0 & 0 & 0 & 0 &   0 &  1,092,867 \\
            & Fixed & Yes & No  &   0 &   0 & 0 & 0 & 0 &   0 & 12,590,366 \\\midrule
            \multirow{5}{*}{267} & BCID  & No  & No  &   0 &   0 & 0 & 0 & 0 &   0 &    202,560 \\
            & BCID  & Yes & No  &   0 &   0 & 0 & 0 & 0 &   0 &     78,079 \\
            & Fixed & No  & No  &   0 &   0 & 0 & 0 & 0 &   0 & 10,218,863 \\
            & Fixed & No  & Yes &   0 &   0 & 0 & 0 & 0 & 150 &  1,092,914 \\
            & Fixed & Yes & No  &   0 &   0 & 0 & 0 & 0 &   0 & 19,300,187 \\
            \bottomrule
    	\end{tabular}
    }
\end{sidewaystable}

\subsubsection{December results}
\label{sec:res_TRIUMF_status_dec}

The status bit results for the December beam test are summarised in table~\ref{tab:proton_stat_dec}. \emph{No} FIFO almost-full or overflow statuses were measured -- this is consistent with the FW counter and HPR out-of-time results described in sections~\ref{sec:res_TRIUMF_testout} and \ref{sec:res_TRIUMF_hpr_dec}, respectively: no ASIC resets were measured.

\begin{sidewaystable}[htbp]
    \centering
	\caption{HPR and register read status bits for December 2020 beam test. The status bit columns indicate the number of packets which had 1 instead of 0 for that status bit. N.B. a packet may count towards the yields in more than one status bit column.}
	\label{tab:proton_stat_dec}
    \resizebox{\textwidth}{!}{
    	\begin{tabular}{clrccccccr} \toprule
            Chip & Idle mode? & Latency & PRFIFO\_almost\_full & LPFIFO\_almost\_full & RegFIFO\_overflow & RegFIFO\_almost\_full & ClusterFIFO\_overflow & ClusterFIFO\_almost\_full & $N_\textrm{packets}$ \\\midrule
            \multirow{2}{*}{259} & No  &  20 & 0 & 0 & 0 & 0 & 0 & 0 &  6,491,530 \\
            & No  & 503 & 0 & 0 & 0 & 0 & 0 & 0 &  2,616,791 \\\midrule
            \multirow{2}{*}{267} & No  &  20 & 0 & 0 & 0 & 0 & 0 & 0 &  4,955,352 \\
            & No  & 503 & 0 & 0 & 0 & 0 & 0 & 0 &  2,639,594 \\\midrule
            \multirow{2}{*}{273} & No  &  20 & 0 & 0 & 0 & 0 & 0 & 0 &  6,910,671 \\
            & No  & 503 & 0 & 0 & 0 & 0 & 0 & 0 &  2,643,497 \\\midrule
            \multirow{3}{*}{274} & Yes &  20 & 0 & 0 & 0 & 0 & 0 & 0 & 13,701,314 \\
            & No  &  20 & 0 & 0 & 0 & 0 & 0 & 0 &  2,075,346 \\
            & No  & 503 & 0 & 0 & 0 & 0 & 0 & 0 &  3,043,142 \\
            \bottomrule
    	\end{tabular}
    }
\end{sidewaystable}

\subsection{SEU cross-section for physics packet clusters}
\label{sec:res_TRIUMF_phys}

Given the L0Buffer is filled in a predictable way, the physics packet clusters returned from the buffer have an expected structure -- the expected structure can be compared to what was actually measured and the differences enumerated.

It is worth noting that there two contributions to the number of the SEUs measured during physics packet readout: the first contribution comes from SEUs accumulated while the data is stored in the L0Buffer and the second contribution comes from SEUs accumulated while the packets are built, processed, and sent. In the subsequent analysis, the two sources are \emph{not} disentangled when calculating the SEU cross-sections.

The SEU cross-section for bit-flips in physics packet clusters, $\sigma_\textrm{SEU}$, is calculated as:

\begin{equation}
    \sigma_\textrm{SEU} = \frac{n_{0\rightarrow1} + n_{1\rightarrow0}}{\int d\phi} \,,
\label{eqn:seu_xs}
\end{equation}

\noindent where $n_{0 \rightarrow 1}$ and $n_{1 \rightarrow 0}$ are the number of $0 \rightarrow 1$  and $1 \rightarrow 0$ bit-flips in the measured clusters, respectively. The cross-section is normalised to the total integrated fluence, $\int d\phi$, accounting for integration time \emph{only} between when the L0Buffer is written to and when it is read out (i.e., when the packet is received). The 95\% confidence interval (assuming Poisson counting statistics) on cross-section, $\Delta_{\sigma_\textrm{SEU}}$, is calculated as:

\begin{equation}
    \Delta_{\sigma_\textrm{SEU}} = 1.96 \times \frac{\sqrt{n_{0\rightarrow1} + n_{1\rightarrow0}}}{\int d\phi} \,.
\label{eqn:seu_cl_xs}
\end{equation}

\noindent In the case of zero measured SEUs, a 95\% upper confidence bound is calculated by assuming 3 SEUs in the cross-section calculation.

\subsubsection{August results}
\label{sec:res_TRIUMF_phys_aug}

For the August beam test, the L0Buffer was filled with ``fixed'' patterns - all 0's or all 1's, alternating every 10~events - or with ``BCID'' patterns - patterns dependent on the BCID. Technical issues were encountered with the BCID fill runs and so physics packets from those runs are not analysed.

Per event, 128~L0A's are sent in order to read out the entire 128-bit deep buffer. Per L0A, 4~reads of the current slice of the L0Buffer are performed. Per read, 1~packet is returned for all 0's and 16~packets for all 1's fills. Thus, per event, 128~groups of either 4 or 64~packets (with each group sharing a common L0ID) are expected for fixed fill runs. Both LP and PR buffer scans are performed, doubling the number of expected physics packets per event.

Within the packets returned for a given L0A, only the last read (i.e., the last 1 or 16~packets in the case all 0's or all 1's fills, respectively) is used in the SEU cross-section calculation. This is because the first, second, and third reads are not independent of the fourth, which also has the longest integration time. In general, the first, second, and third reads were measured for posterity and only scrutinised in select cases.\footnote{For an example of this, in situations where the fourth read is truncated due to edge effects, the third read is analysed in place of the fourth. However, this is of no consequence with respect to the measured results.}

The measured and expected clusters are converted to their corresponding 256-bit strip data sequences, and the measured and expected sequences are compared to one another and the number of bit-flips are counted. The LP and PR buffer scans are combined when counting the number of bit-flips. The SEU cross-sections for bit-flips in physics packet clusters are summarised in table~\ref{tab:proton_phys_xs_aug}.

\begin{table}[htbp]
    \centering
    \caption{A summary of the SEU cross-sections for bit-flips in physics packet clusters per chip per running mode for the August 2020 beam test.}
    \label{tab:proton_phys_xs_aug}
    \resizebox{\textwidth}{!}{
        \begin{tabular}{cllrrrrr}
        \toprule
           Chip                 & Glitch filter? & Clock disabled? & $n_{0\rightarrow1}$ & $n_{1\rightarrow0}$ & $N_\textrm{bits,total}$ & $\int d\phi$ [p/\cmsq] & $\sigma_{\mathrm{SEU}} \pm \Delta_{\sigma_\textrm{SEU}}$ [\cmsq/p] \\
        \midrule
           002                  & No             & No                  &              33,699 &              34,546 &              15,250,176 &   $1.86 \times 10^{16}$ &                                   $(3.69 \pm 0.03) \times 10^{-12}$ \\
        \midrule
           \multirow{3}{*}{259} & No             & No                  &              16,343 &              16,506 &               7,642,368 &   $4.47 \times 10^{15}$ &                                   $(7.35 \pm 0.08) \times 10^{-12}$ \\
                                & No             & Yes                 &               1,016 &                 784 &                 411,136 &   $2.34 \times 10^{14}$ &                                   $(7.7  \pm 0.4)  \times 10^{-12}$ \\
                                & Yes            & No                  &              13,282 &              13,913 &               6,321,408 &   $3.83 \times 10^{15}$ &                                   $(7.09 \pm 0.08) \times 10^{-12}$ \\
        \midrule
           \multirow{3}{*}{261} & No             & No                  &              18,679 &              18,853 &               8,543,488 &   $6.67 \times 10^{15}$ &                                   $(5.63 \pm 0.05) \times 10^{-12}$ \\
                                & No             & Yes                 &               2,233 &               1,971 &                 930,304 &   $6.38 \times 10^{14}$ &                                   $(6.6  \pm 0.2)  \times 10^{-12}$ \\
                                & Yes            & No                  &              18,848 &              19,643 &               8,687,360 &   $6.20 \times 10^{15}$ &                                   $(6.21 \pm 0.06) \times 10^{-12}$ \\
        \midrule
           \multirow{3}{*}{267} & No             & No                  &               8,459 &               8,775 &               4,129,280 &   $2.85 \times 10^{15}$ &                                   $(6.04 \pm 0.09) \times 10^{-12}$ \\
                                & No             & Yes                 &               1,342 &               1,053 &                 562,432 &   $3.38 \times 10^{14}$ &                                   $(7.1  \pm 0.3)  \times 10^{-12}$ \\
                                & Yes            & No                  &              15,235 &              14,798 &               7,214,592 &   $5.04 \times 10^{15}$ &                                   $(5.97 \pm 0.07) \times 10^{-12}$ \\
        \midrule
           All V1 & \multicolumn{2}{l}{\multirow{2}{*}{All modes}} & \multirow{2}{*}{129,136} & \multirow{2}{*}{130,842} & \multirow{2}{*}{59,692,544} & \multirow{2}{*}{$4.89 \times 10^{16}$} & \multirow{2}{*}{$(5.32 \pm 0.02) \times 10^{-12}$} \\
           chips & & & & & & \\
        \bottomrule
        \end{tabular}
    }
\end{table}

Within a particular chip, the different running scenarios (i.e., glitch filter enabled/disabled, clock enabled/disabled) result in cross-sections which are very consistent with one another. This behaviour is expected -- the glitch filter counts glitches on the input clock which is only expected when the generation is part of the system under test, as is the case with a Hybrid Control Chip (HCC)~\cite{TDRs}. Moreover, disabling of the clock only enhances the number of SEUs in triplicated registers (see section~\ref{sec:res_TRIUMF_reg}).

Additionally, the cross-sections obtained for different chips are very consistent with one another, all $\mathcal{O}(10^{-12})$~\cmsq/p. The V0 and V1 chips measure cross-sections of $(3.69 \pm 0.03) \times 10^{-12}$~\cmsq/p and $(5.32 \pm 0.02) \times 10^{-12}$~\cmsq/p, respectively. While the two cross-sections don't agree within statistical error, they are likely much more consistent when including systematic effects such as the uncertainty on the integrated fluence.\footnote{There are two sources of uncertainties affecting the integrated fluence: the uncertainty on the number of SEM counts, which has a linear effect, and the uncertainty on the distance of the SCBs from the end of the beampipe, which has a quadratic effect. The number of SEM counts comes with a Poisson counting uncertainty, which is expected to be small, and an uncertainty on the efficiency for measuring a particular SEM count, which is unknown. Focusing on the uncertainty on the distance, if we made the (\emph{very}) conservative assumption of a 25\% uncertainty, then this translates to a 50\% uncertainty on the integrated fluence. Even still, this does not affect our conclusion concerning the effect of SEUs in physics packet clusters on the detector's performance.}

Very rarely, the output data was truncated during readout as a result of issues encountered due to edge effects in the DAQ software -- excluding these cases, no SEUs were measured as part of packet building, processing, and sending. This is expected, as these mechanisms utilise TMR protection.

\subsubsection{December results}
\label{sec:res_TRIUMF_phys_dec}

During the December 2020 beam test, the L0Buffer was repeatedly filled and read out (\emph{only} LP buffer scans) in groups of 8~triggers, with the fill pattern set according to the BCID. To represent a more realistic running scenario, the L0Buffer was filled and immediately read out for each group of 8~triggers. The measured packets were grouped by closest-in-time (i.e., within 1000~BCs of one another). The integration time for the measured packets is assumed to be constant and given by:

\begin{equation}
    \Delta t = (40\,\textrm{MHz})^{-1} \times (255 \times (7 - i) + \textrm{latency}) \,,
\end{equation}

\noindent where $i$ is the index of the packet in the group of 8 ($i = 0,\ldots,7$) and the latency is either 20 or 503, depending on the run. N.B. the integration times used in December were \emph{considerably} shorter than those used in August: $\mathcal{O}(10)$~$\upmu$s/packet for December versus $\mathcal{O}(10)$~s/packet for August. However, this difference is accounted for in the total integrated fluence.

With the BCID-dependent fills, the clusters returned in a packet are predictable based on the BCID. The BCID decrements by 1 from one packet to the next within a group (accounting for overflow, modulo 256). On group edges, the BCID increments by the number of triggers sent, 8, plus the number of measured packets in the current group, 8 (accounting for overflow, modulo 256). Packets are identified as ``bad'' if the BCID corresponding to the returned clusters does not match the value expected within the larger pattern or if the returned clusters do not correspond to any of the clusters possible with BCID fills.

The number of bit-flips is calculated in the same way as for the August results (i.e., by comparing measured and expected strip data sequences). The SEU cross-sections for bit-flips in physics packet clusters are summarised in table~\ref{tab:proton_phys_xs_dec}.

\begin{table}[htbp]
    \centering
    \caption{A summary of the SEU cross-sections for bit-flips in physics packet clusters per chip per running mode for the December 2020 beam test. Cross-sections with a prefixed asterisk (*) indicate 95\% upper confidence bounds assuming 3 SEUs.}
    \label{tab:proton_phys_xs_dec}
    \resizebox{\textwidth}{!}{
        \begin{tabular}{clrrrrr}
        \toprule
           Chip                 &   Latency & $n_{0\rightarrow1}$ & $n_{1\rightarrow0}$ & $N_\textrm{bits,total}$ & $\int d\phi$ [p/\cmsq] & $\sigma_{\mathrm{SEU}} \pm \Delta_{\sigma_\textrm{SEU}}$ [\cmsq/p] \\
        \midrule
           \multirow{2}{*}{259} &        20 &                   0 &                   0 &             219,809,024 &   $2.09 \times 10^{10}$ &                                             *$1.44 \times 10^{-10}$ \\
                                &       503 &                   0 &                   0 &             299,576,832 &   $4.44 \times 10^{10}$ &                                             *$6.78 \times 10^{-11}$ \\
        \midrule
           \multirow{2}{*}{267} &        20 &                   1 &                   0 &             218,816,256 &   $1.73 \times 10^{10}$ &                                     $(0.6 \pm 1.1) \times 10^{-10}$ \\
                                &       503 &                   1 &                   0 &             299,680,256 &   $3.70 \times 10^{10}$ &                                         $(3 \pm 5) \times 10^{-11}$ \\
        \midrule
           \multirow{2}{*}{273} &        20 &                   2 &                   1 &             826,031,104 &   $8.05 \times 10^{10}$ &                                         $(4 \pm 4) \times 10^{-11}$ \\
                                &       503 &                   0 &                   0 &             300,588,544 &   $4.28 \times 10^{10}$ &                                             *$7.02 \times 10^{-11}$ \\
        \midrule
           \multirow{2}{*}{274} &        20 &                   0 &                   0 &             259,894,528 &   $2.25 \times 10^{10}$ &                                             *$1.34 \times 10^{-10}$ \\
                                &       503 &                   0 &                   0 &             349,664,768 &   $4.41 \times 10^{10}$ &                                             *$6.81 \times 10^{-11}$ \\
        \midrule
           All chips            & 20 \& 503 &                   4 &                   1 &           2,774,061,312 &   $3.10 \times 10^{11}$ &                                     $(1.6 \pm 1.4) \times 10^{-11}$ \\
        \bottomrule
        \end{tabular}
    }
\end{table}

Summing all chips, the SEU cross-section for bit-flips in physics packet clusters is $(1.6 \pm 1.4) \times 10^{-11}$~\cmsq/p. This result is consistent with the cross-section measured in August for the V1 ASICs, $(5.32 \pm 0.02) \times 10^{-12}$~\cmsq/p. Additionally, for the run scenarios where no bit-flips were observed, the upper confidence bounds of $\mathcal{O}(10^{-11})$ to $\mathcal{O}(10^{-10})$~\cmsq/p are nearly consistent with the cross-sections for the runs where bit-flips were observed in both August and December.

As with the August results, DAQ-related issues sometimes caused only 7~triggers to be sent instead of 8. However, no SEUs were measured as part of packet building, processing, and sending, as expected.

\subsubsection{Combined results and estimated impact on operation}
\label{sec:res_TRIUMF_phys_comb}

Altogether, the combined cross-section for V1 chips using data from August and December is $(5.32 \pm 0.02) \times 10^{-12}$~\cmsq/p. This combined result is dominated by the August runs due to their much larger integrated fluence as compared to those in December. As the time in the pipeline is long for (i.e., \unit[$\mathcal{O}(10)$]{s}) for the data measured in August and comparatively short (i.e., \unit[$\mathcal{O}(10)$]{$\upmu$s}) for the data measured in December, the combined number of SEUs is dominated by those accumulated while the data is in the L0Buffer.

We can estimate the number of hit errors per second during typical ITk operation at the HL-LHC. ATLAS simulations show worse-case expected fluences of $\mathcal{O}(10^{-3})$~hadrons/\cmsq/$pp$~collision (where $pp$ collision refers to a proton-proton collision). Assuming a pileup of $\langle\mu\rangle = 200$ and given the rate of BCs is \unit[40]{MHz} (i.e., the number of $pp$ collisions per second)\footnote{The collision rate of \unit[40]{MHz} represents a worst-case scenario, as approximately only 2,700 in 3,500~BCs are filled.}, the hadron flux, $\Phi_\textrm{hadrons}$, is given by:

\begin{equation}
    \label{eqn:flux}
    \begin{split}
        \Phi_\textrm{hadrons} & = \mathcal{O}(10^{-3})~\textrm{hadrons/\cmsq/$pp$ collision} \times 200 \\
        & \,\,\,\,\,\,\,\, \times (40\times10^{6})~\textrm{$pp$ collisions/s} \\
        & = \mathcal{O}(10^7)~\textrm{hadrons/\cmsq/s} \,.
    \end{split}
\end{equation}

\noindent The data acquisition scheme in December is a more realistic running scenario (i.e., the event buffer is filled and rapidly read-out). For December, the average time in the pipeline for data is \unit[$\mathcal{O}(10)$]{$\upmu$s}. Additional simulations show the worst case readout rates in the ITk are $10/9 \approx 1.1$~packets/event/ABCStar, where an event is defined as a \emph{readout} event. Given equation~\ref{eqn:flux} and the above information, the rate of hit errors due to SEUs in physics packet clusters, $\textrm{Rate}_\textrm{SEU}$, is given by:

\begin{equation}
    \begin{split}
        \textrm{Rate}_\textrm{SEU} & = \left( \Phi_\textrm{hadrons} \times \frac{\langle\textrm{time in pipeline}\rangle}{\textrm{packet}} \right) \times \frac{\textrm{packets}}{\textrm{event}} \times \sigma_\textrm{SEU} \\
        & = (\mathcal{O}(10^7)~\textrm{hadrons/\cmsq/s} \times \mathcal{O}(10^{-5})~\textrm{s/packet}) \\
        & \,\,\,\,\,\,\,\, \times 1.1~\textrm{packets/event} \times \mathcal{O}(10^{-12})~\textrm{\cmsq/p/ABCStar} \\
        & = \mathcal{O}(10^{-10})~\textrm{errors/event/ABCStar} \,.
    \end{split}
\end{equation}

\noindent Here, the bracketed factors correspond to the effective integrated fluence per packet. So we expect $\mathcal{O}(10^{-10})$~hit errors per readout event per ABCStar. Given there are $\sim\!230,000$ ABCStar ASICs in the ITk strip tracker\footnote{The number of ASICs also represents a worst-case scenario, as only a subset of all ASICs will be read out during a collision.} and assuming a trigger rate of \unit[1]{MHz}, this equates to $\mathcal{O}(20)$~hit errors/s during normal operation. This error rate is \emph{significantly} below the noise occupancy due thermal noise expected for the detector (i.e., $10^{-2}$ to $10^{-1}$~hit errors per readout event per ABCStar) and so is of no practical concern.

\subsection{SEU cross-section for register read data}
\label{sec:res_TRIUMF_reg}

A measurement of the SEU cross-section is calculated using the register reads performed during both proton beam tests. When reading a register, the returned packet is required to have a structure consistent with that described in table~\ref{tab:packets}.

As SEUs are identified by comparing an expected payload to a measured one, the only register types considered are those whose content is well defined: SCReg, ADCS, MaskInput, CREG, STAT (\emph{only} the fuse ID register, \texttt{0x32}), TrimDAC, and CalREG (as a reminder, the different register types are enumerated in table~\ref{tab:registers}). The HitCountREG registers, which summarise front-end cluster information, are (implicitly) analysed as part of the physics data cross-section analysis (see section~\ref{sec:res_TRIUMF_phys}. The HPR register, which contains the information returned by HPR packets, is reserved for the HPR packet analysis (see section~\ref{sec:res_TRIUMF_hpr}).

The SEU cross-section for bit-flips in the register read data, $\sigma_{\mathrm{SEU}}$, and its associated 95\% confidence interval, $\Delta_{\sigma_{\mathrm{SEU}}}$, are calculated using equations~\ref{eqn:seu_xs} and \ref{eqn:seu_cl_xs}, respectively. The integrated fluence is only counted between the times when a register is written to and when it is read out.

\subsubsection{August results}
\label{sec:res_TRIUMF_reg_aug}

In August, the ADCS registers' contents are static in time. This is also true for the fuse ID STAT register. The MaskInput and TrimDAC registers are filled with all 1's for 10 events followed by all 0's for 10 events (where an event consists of a full read out of all registers), and this pattern repeats for the duration of all runs. The CalREG registers are filled with all 0's for 10 events followed by all 1's for 10 events (this is similarly true for the HitCountREG registers for fixed fill runs). For the V0 chip, the CREG registers are static in time; for the V1 chips, only the CREG register \texttt{0x20} is static in time (the other CREG register, \texttt{0x21}, is dynamic with unobvious patterns). As a result, only \texttt{0x20} is included as part of the CREG registers for V1 chips. It's important to note that the SCReg was \emph{not} measured as part of the August beam test due a mistake in the measurement software. In total, 70 and 61 registers were included in the SEU cross-section calculation for the V0 and V1 chips, respectively.

\begin{figure}[htbp]
    \centering
    \includegraphics[width=1.\linewidth]{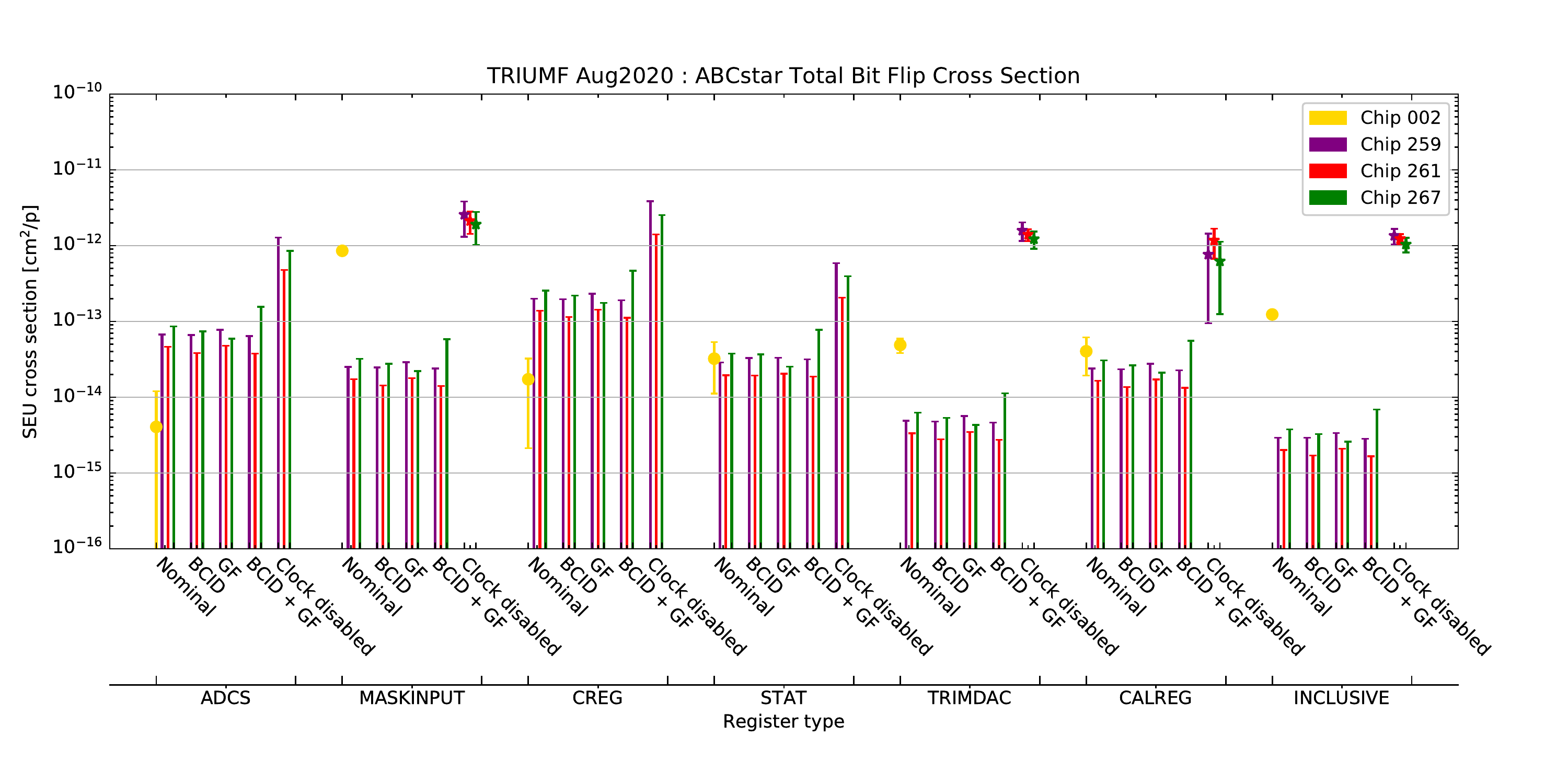}
    \caption{The SEU cross-sections for bit-flips in register read data for the August 2020 beam test. Visible data points as well as their errors are calculated using equations~\ref{eqn:seu_xs} and \ref{eqn:seu_cl_xs}. Data points which are not visible on the scale of the graph correspond to 95\% upper confidence bounds assuming 3 SEUs. ``Nominal'', ``BCID'' (BCID fills), ``GF'' (glitch filter enabled),  ``BCID + GF'' (BCID fills with glitch filter enabled), and ``Clock disabled'' refer to the different running scenarios. ``INCLUSIVE'' is inclusive in the other register types. N.B. the V0 chip, 002, was operated only in the nominal running mode.}
    \label{fig:proton_reg_xs_aug}
\end{figure}

Figure~\ref{fig:proton_reg_xs_aug} shows the SEU cross-sections for bit-flips in register read data as a function of register type, running mode, and chip. Table~\ref{tab:proton_reg_xs_aug} summarises the inclusive cross-sections for each chip.

\begin{table}[htbp]
    \centering
    \caption{A summary of the inclusive SEU cross-sections for bit-flips in register read data per chip per running mode for the August 2020 beam test. Cross-sections with a prefixed asterisk (*) indicate 95\% upper confidence bounds assuming 3 SEUs.}
    \label{tab:proton_reg_xs_aug}
    \resizebox{\textwidth}{!}{
        \begin{tabular}{clllrrrrr}
            \toprule
            Chip & Fill type & Glitch filter? & Clock disabled? & $n_{0\rightarrow1}$ & $n_{1\rightarrow0}$ & $N_\textrm{bits,total}$ & $\int d\phi$ [p/\cmsq] & $\sigma_{\mathrm{SEU}} \pm \Delta_{\sigma_\textrm{SEU}}$ [\cmsq/p] \\ \midrule
            002 & Fixed & No  & No  & 16 & 378 & 3,518,912 & $3.19 \times 10^{15}$ & $(1.24 \pm 0.12) \times 10^{-13}$ \\ \midrule
            \multirow{5}{*}{259} & Fixed & No  & No  &  0 &   0 & 1,715,456 & $1.02 \times 10^{15}$ & *$2.93 \times 10^{-15}$ \\
            & BCID  & No  & No  &  0 &   0 & 1,661,856 & $1.03 \times 10^{15}$ & *$2.91 \times 10^{-15}$ \\
            & Fixed & Yes & No  &  0 &   0 & 1,452,128 & $8.87 \times 10^{14}$ & *$3.39 \times 10^{-15}$ \\
            & BCID  & Yes & No  &  0 &   0 & 1,669,376 & $1.06 \times 10^{15}$ & *$2.83 \times 10^{-15}$ \\
            & Fixed & No  & Yes &  3 &  69 &    68,128 & $5.32 \times 10^{13}$ & $(1.4 \pm 0.3) \times 10^{-12}$ \\ \midrule
            \multirow{5}{*}{261} & Fixed & No  & No  &  0 &   0 & 1,458,240 & $1.49 \times 10^{15}$ & *$2.01 \times 10^{-15}$ \\
            & BCID  & No  & No  &  0 &   0 & 1,735,744 & $1.77 \times 10^{15}$ & *$1.70 \times 10^{-15}$ \\
            & Fixed & Yes & No  &  0 &   0 & 1,477,696 & $1.44 \times 10^{15}$ & *$2.09 \times 10^{-15}$ \\
            & BCID  & Yes & No  &  0 &   0 & 1,888,000 & $1.81 \times 10^{15}$ & *$1.66 \times 10^{-15}$ \\
            & Fixed & No  & Yes & 12 & 166 &   126,624 & $1.44 \times 10^{14}$ & $(1.23 \pm 0.18) \times 10^{-12}$ \\ \midrule
            \multirow{5}{*}{267} & Fixed & No  & No  &  0 &   0 & 1,212,608 & $7.97 \times 10^{14}$ & *$3.78 \times 10^{-15}$ \\
            & BCID  & No  & No  &  0 &   0 & 1,884,128 & $9.19 \times 10^{14}$ & *$3.27 \times 10^{-15}$ \\
            & Fixed & Yes & No  &  0 &   0 & 2,264,416 & $1.16 \times 10^{15}$ & *$2.59 \times 10^{-15}$ \\
            & BCID  & Yes & No  &  0 &   0 &   725,504 & $4.36 \times 10^{14}$ & *$6.87 \times 10^{-15}$ \\
            & Fixed & No  & Yes &  3 &  79 &   124,864 & $7.88 \times 10^{13}$ & $(1.0 \pm 0.2) \times 10^{-12}$ \\ \midrule
            All V1 & \multicolumn{3}{l}{\multirow{2}{*}{All modes except clock disabled}} & \multirow{2}{*}{0} & \multirow{2}{*}{0} & \multirow{2}{*}{19,145,152} & \multirow{2}{*}{$1.39 \times 10^{16}$} & \multirow{2}{*}{*$2.16 \times 10^{-16}$} \\
            chips & & & & & & \\ \bottomrule
        \end{tabular}
    }
\end{table}

The effect of the register triplication is striking: while the V0 ASICs saw a nonzero number of SEUs,\footnote{The asymmetry in the number of $0\rightarrow1$ (16) and $1\rightarrow0$ (378) bit flips measured from the V0 register read packets is a result of a vulnerability in the V0 ASICs which was corrected in the V1 ASICs.} the V1 ASICs saw \emph{zero} SEUs for $\mathcal{O}(10^{7})$ bits measured. By disabling the clock of the triplicated registers in the V1 ASICs, an enhancement of the SEU cross-section is measured, as expected.\footnote{The V1 register read SEU cross-section with the clock disabled is not necessarily comparable to the V0 register read SEU cross-section. Rather, this measurement is simply a check to verify that the triplication is indeed working in the V1 ASICs.} Additionally, there is no difference in the behaviour between the fixed and BCID fill running modes as well as the glitch filter disabled and enabled running modes. This is expected, as the fixed and BCID fill running modes affect the clusters measured on front-end of the chip, and the glitch filter counters glitches on the \unit[40]{MHz} clock input pad, which, for the purposes of the SEU beam test, was an ideal source.

\subsubsection{December results}
\label{sec:res_TRIUMF_reg_dec}

In December, the SCReg, ADCS, MaskInput, CREG, STAT (fuse ID only), TrimDAC, and CalREG registers are all static in time. Additionally, with the exception of the fuse ID register, all of the above registers are written to \emph{once} at the start of a run. Accordingly, the SEU calculation verifies that contents of each register is constant with time and integrates the fluence from the initial write time of all registers to their last respective read time. In total, 63 registers were included in the SEU cross-section calculation.

\begin{figure}[htbp]
    \centering
    \includegraphics[width=1.\linewidth]{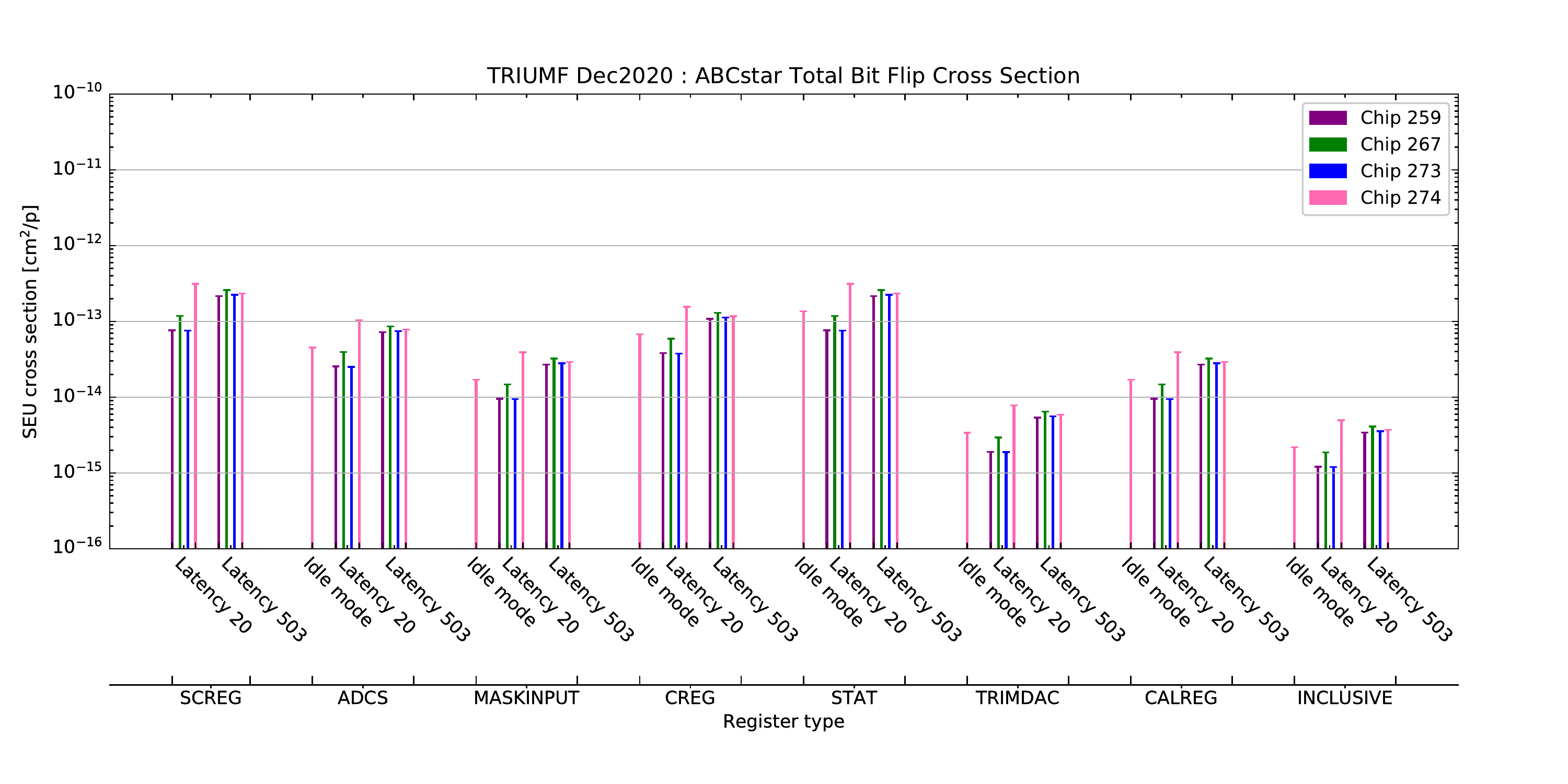}
    \caption{The SEU cross-sections for bit-flips in register read data for the December 2020 beam test. Visible data points as well as their errors are calculated using equations~\ref{eqn:seu_xs} and \ref{eqn:seu_cl_xs}. Data points which are not visible on the scale of the graph correspond to 95\% upper confidence bounds assuming 3 SEUs. ``Idle mode'', ``Latency 20'', and ``Latency 503'' refer to the different running scenarios. ``INCLUSIVE'' is inclusive in the other register types. N.B. for Chip 274's idle mode runs, register \texttt{0x00} (``SCREG'') was not measured.}
    \label{fig:proton_reg_xs_dec}
\end{figure}

Figure~\ref{fig:proton_reg_xs_dec} shows the SEU cross-sections for bit-flips in register read data as a function of register type, running mode, and chip. Table~\ref{tab:proton_reg_xs_dec} summarises the inclusive cross-sections for each chip.

\begin{table}[htbp]
    \centering
    \caption{A summary of the inclusive SEU cross-sections for bit-flips in register read data per chip per running mode for the December 2020 beam test. Cross-sections with a prefixed asterisk (*) indicate 95\% upper confidence bounds assuming 3 SEUs.}
    \label{tab:proton_reg_xs_dec}
    \resizebox{\textwidth}{!}{
        \begin{tabular}{clrrrrrr}
            \toprule
            Chip & Idle mode? & Latency & $n_{0\rightarrow1}$ & $n_{1\rightarrow0}$ & $N_\textrm{bits,total}$ & $\int d\phi$ [p/\cmsq] & $\sigma_{\mathrm{SEU}} \pm \Delta_{\sigma_\textrm{SEU}}$ [\cmsq/p] \\ \midrule
            \multirow{2}{*}{259} & No  &  20 & 0 & 0 &  7,638,624 & $2.48 \times 10^{15}$ & *$1.21 \times 10^{-15}$ \\
            & No  & 503 & 0 & 0 &  2,941,344 & $8.76 \times 10^{14}$ & *$3.42 \times 10^{-15}$ \\ \midrule
            \multirow{2}{*}{267} & No  &  20 & 0 & 0 &  5,695,200 & $1.60 \times 10^{15}$ & *$1.88 \times 10^{-15}$ \\
            & No  & 503 & 0 & 0 &  2,941,344 & $7.30 \times 10^{14}$ & *$4.11 \times 10^{-15}$ \\ \midrule
            \multirow{2}{*}{273} & No  &  20 & 0 & 0 &  8,116,416 & $2.50 \times 10^{15}$ & *$1.20 \times 10^{-15}$ \\
            & No  & 503 & 0 & 0 &  2,949,408 & $8.42 \times 10^{14}$ & *$3.57 \times 10^{-15}$ \\ \midrule
            \multirow{3}{*}{274} & Yes &  20 & 0 & 0 & 23,675,072 & $1.37 \times 10^{15}$ & *$2.19 \times 10^{-15}$ \\
            & No  &  20 & 0 & 0 &  2,556,288 & $6.04 \times 10^{14}$ & *$4.95 \times 10^{-15}$ \\
            & No  & 503 & 0 & 0 &  3,429,216 & $8.07 \times 10^{14}$ & *$3.72 \times 10^{-15}$ \\ \midrule
            All chips & \multicolumn{2}{l}{All modes} & 0 & 0 & 59,942,912 & $1.18 \times 10^{16}$ & *$2.54 \times 10^{-16}$ \\ \bottomrule
        \end{tabular}
    }
\end{table}

As with the V1 measurements from August, \emph{zero} SEUs were observed for $\mathcal{O}(10^{7})$ bits measured. The upper limits for each register type are very consistent between the two beam test campaigns. This is not necessarily expected for upper limits in general, but it indicates a comparable amount of data was obtained by each campaign, which is expected. Combining the inclusive data for all V1 chips (excluding the runs with the clock disabled) in August and December, we have 95\% upper confidence bound of $1.17 \times 10^{-16}$~\cmsq/p on the SEU cross-section for bit-flips in register read data. Altogether, the results indicate that the triplication implemented in registers of V1 ABCStar ASICs offers excellent protection against SEUs in those registers. This performance is well beyond the needs of operation.

\subsection{Current measurements}
\label{sec:res_TRIUMF_curr_meas}

\begin{figure}[htbp]
    \centering
    \begin{subfigure}{.49\textwidth}
        \centering
        \includegraphics[width=1.\linewidth]{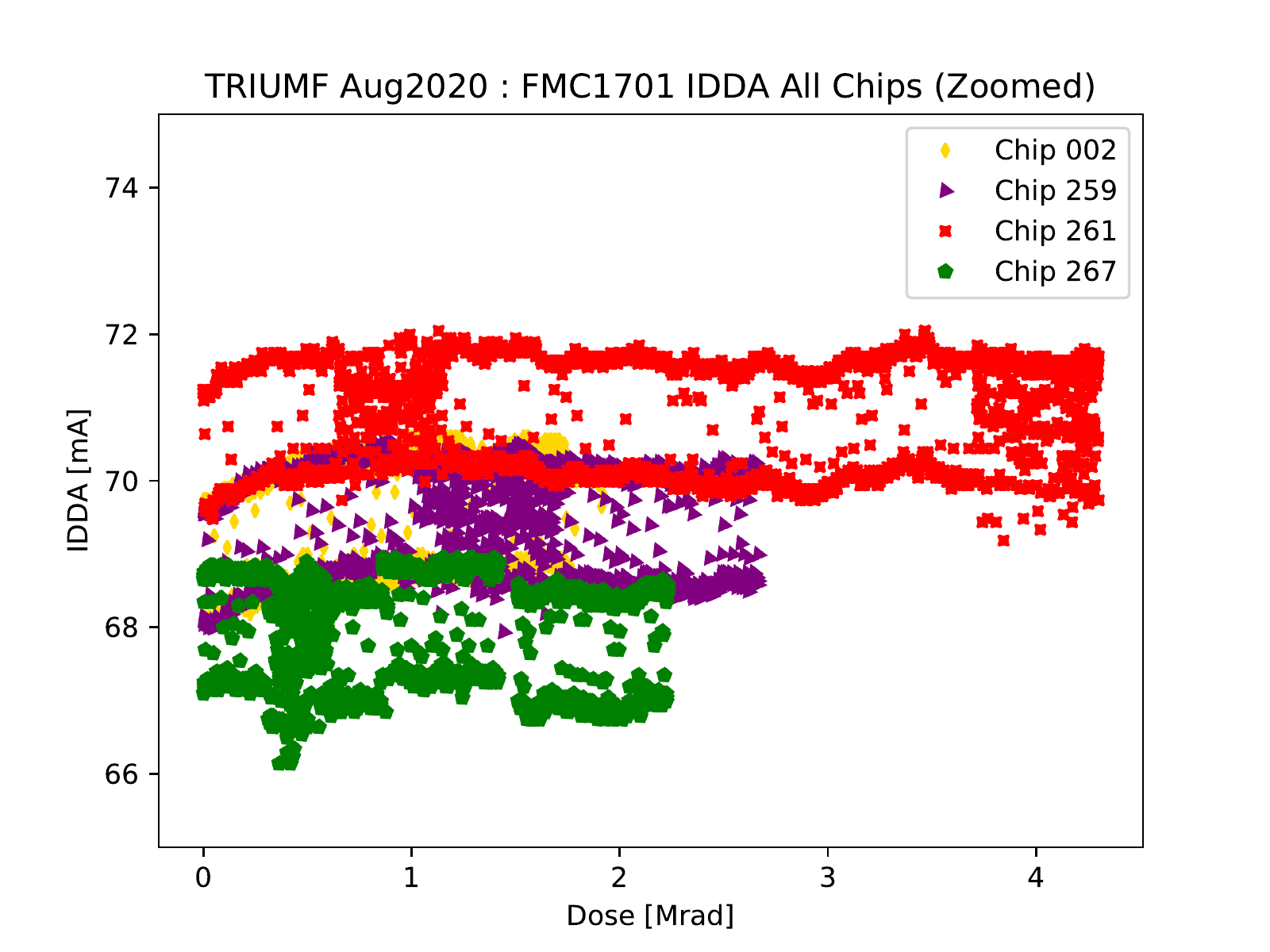}
        \caption{Analogue current}
        \label{fig:proton_analog_curr_aug}
    \end{subfigure}
    \hfill
    \centering
    \begin{subfigure}{.49\textwidth}
        \centering
        \includegraphics[width=1.\linewidth]{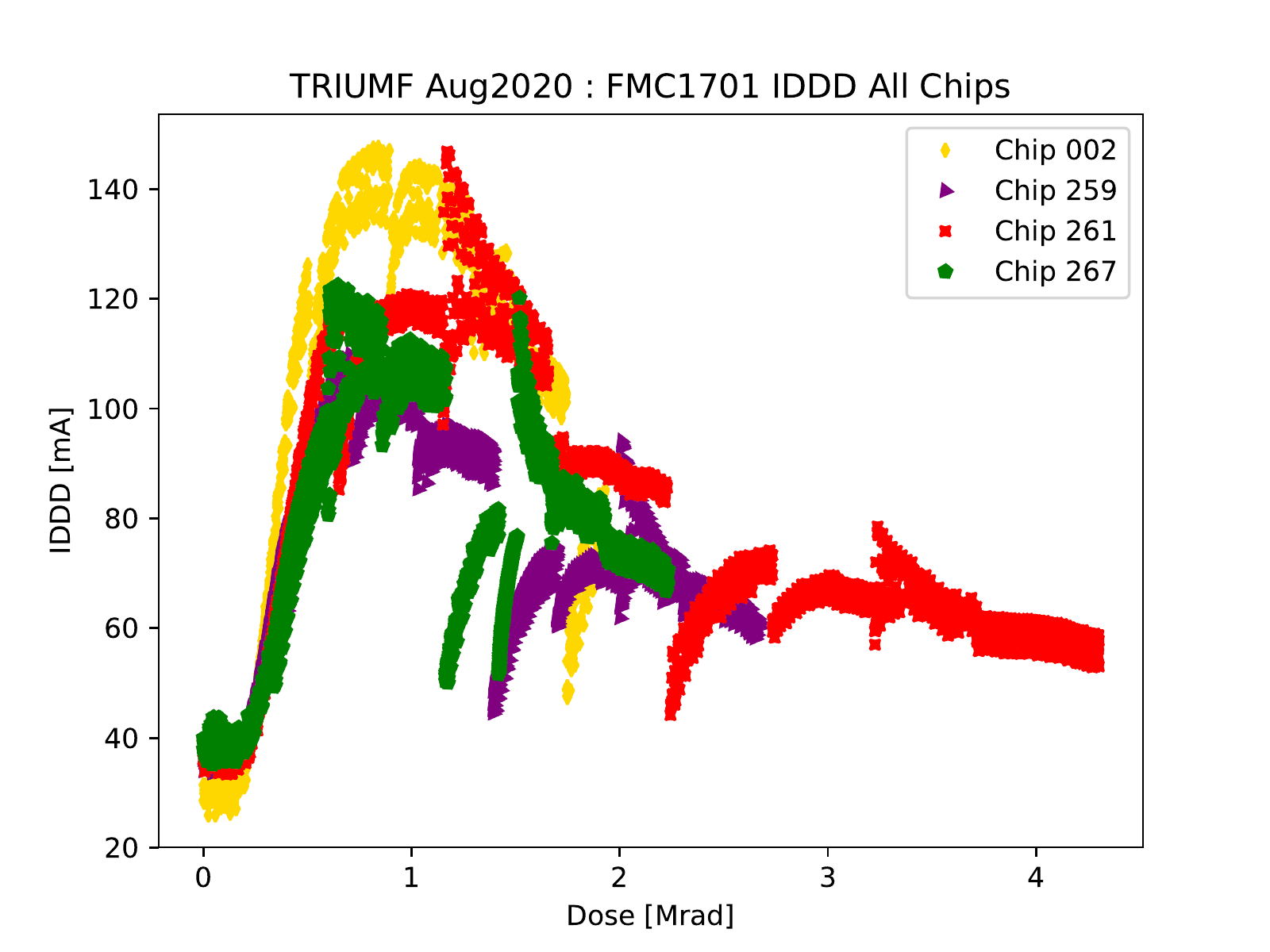}
        \caption{Digital current}
        \label{fig:proton_digital_curr_aug}
    \end{subfigure}
    \caption{The (a) analogue and (b) digital currents drawn by each ABCStar ASIC as measured by the FMC-1701 during August 2020 beam test. The currents are shown as a function of dose. The analogue currents for the yellow data points lie nearly underneath the purple data points. The piece-wise behaviour of the digital current curves is a result of stopping/starting data collection runs. N.B. the band-like structures in the analogue currents are a result of switching between the different L0Buffer fill configurations during the beam test.}
    \label{fig:proton_curr_aug}
\end{figure}

\begin{figure}[htbp]
    \centering
    \begin{subfigure}{.49\textwidth}
        \centering
        \includegraphics[width=1.\linewidth]{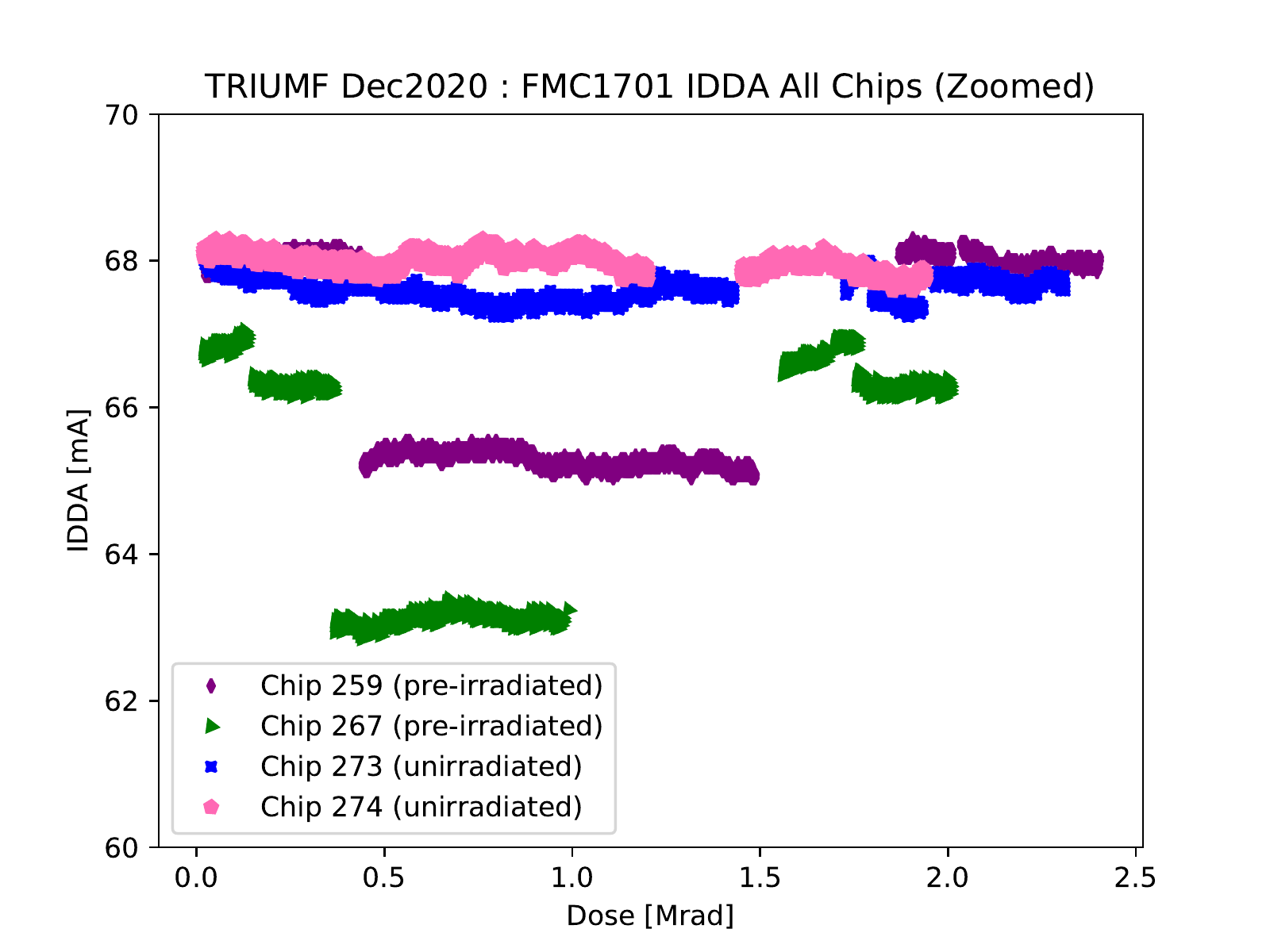}
        \caption{Analogue current}
        \label{fig:proton_analog_curr_dec}
    \end{subfigure}
    \hfill
    \centering
    \begin{subfigure}{.49\textwidth}
        \centering
        \includegraphics[width=1.\linewidth]{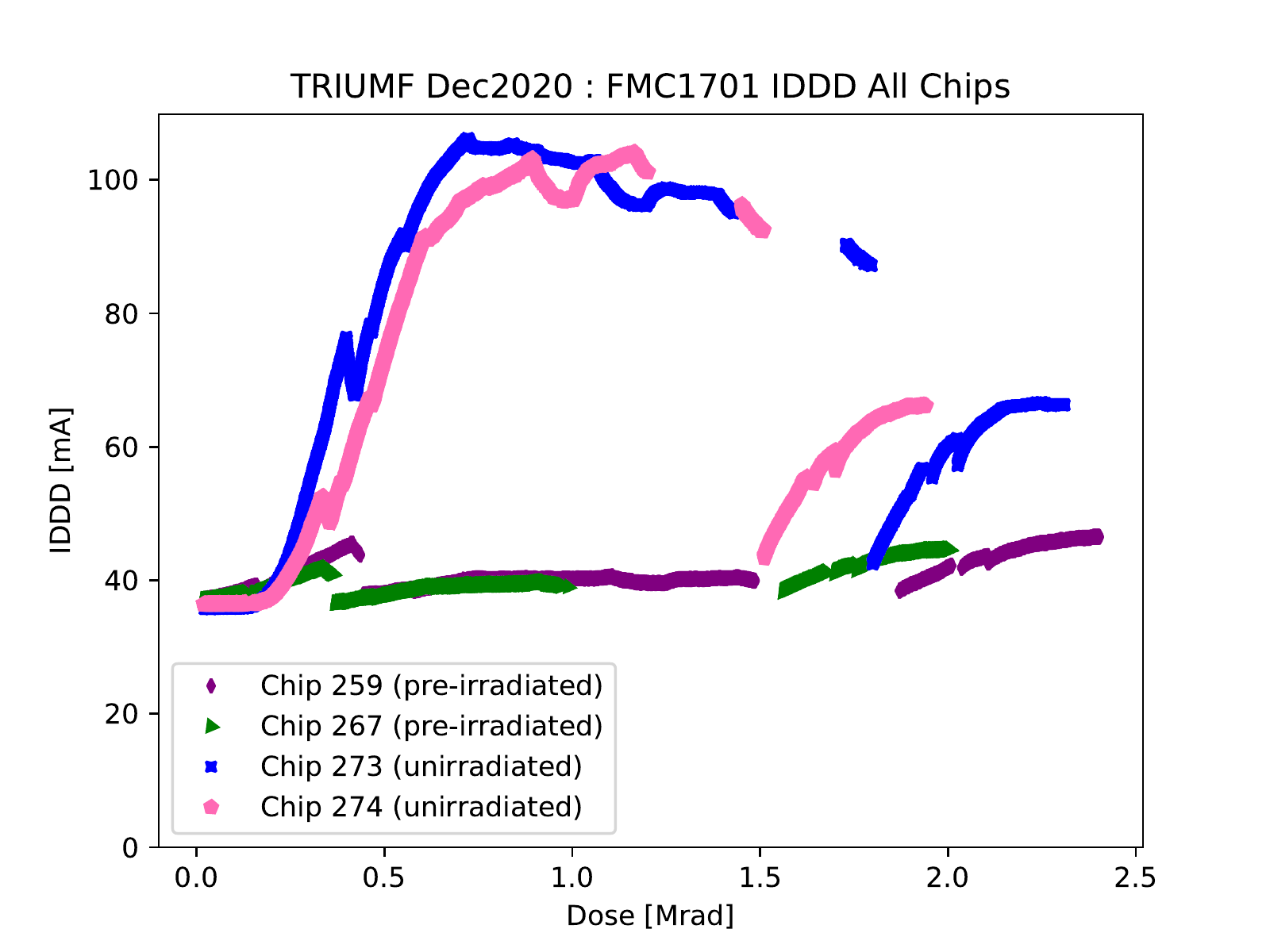}
        \caption{Digital current}
        \label{fig:proton_digital_curr_dec}
    \end{subfigure}
    \caption{The (a) analogue and (b) digital currents drawn by each ABCStar ASIC as measured by the FMC-1701 during December 2020 beam test. The currents are shown as a function of dose. The piece-wise behaviour of the digital current curves is a result of stopping/starting data collection runs. Two new ASICs were tested and compared with two previously irradiated ASICs, confirming that the total current increase during irradiation is lower for pre-irradiated ASICs of the ABCStar V1 generation.}
    \label{fig:proton_curr_dec}
\end{figure}

The analogue and digital currents measured by the FMC-1701 on the SCB for all chips from the August and December beam tests are shown in figures~\ref{fig:proton_curr_aug} and \ref{fig:proton_curr_dec}, respectively.

The analogue currents remain constant at approximately \unit[70]{mA}, regardless of the radiation dose delivered to the ASIC. This is consistent with what was observed for ABC130 ASICs~\cite{ABC130}. The digital currents see an initial increase in current, reaching a maximum of \unit[140]{mA} at \unit[1]{Mrad} before decreasing to \unit[60]{mA} at higher doses. This TID bump has been studied extensively in ATLAS ASICs (see for instance section~4.10 of reference~\cite{ABC130}). Pre-irradiation digital currents of \unit[40]{mA} followed by a near 100\% increase in current near \unit[1]{Mrad} before returning to nominal levels is consistent with what was observed for ABC130 ASICs. Given that the ABCStar ASICs were manufactured with the same technology as the ABC130 ASICs, this indicates the TID bump to be a stable feature. It should be noted that the digital current increase is roughly comparable for the ABCStar V0 and V1 ASICs.

During the December beam test, where two new ABCStar V1 ASICs (273 and 274) were irradiated alongside two previously tested ASICs (259 and 267), a comparable current measurement shows a similar increase of the digital current for the new ASICs, as shown in figure~\ref{fig:proton_digital_curr_dec}). The two previously irradiated ASICs show a digital current increase much lower than the unirradiated ASICs, confirming the effect of pre-irradiation on the TID bump observed for ABC130 ASICs~\cite{ABC130}.

\FloatBarrier

\section{Complementary results from heavy ion measurements}
\label{sec:hi}

Proton interactions within the ASIC volume cause the protons to recoil, and the deposited energy depends on their recoil angle. Therefore, despite a uniform beam energy of \unit[480]{MeV}, the observed SEUs correspond to an integrated deposited energy spectrum. In order to calculate the expected rate of SEUs within the detector based on the expected particle energies, complementary measurements were performed for a single ABCStar V1 ASIC using heavy ions (an overview of the ion energies used here is given in table~\ref{tab:LET}) at the cyclotron facility at Louvain-la-Neuve, Belgium~\cite{UCL}, where the deposited energy corresponded to the ion energy.

\begin{table}[htbp]
	\centering
	\small
	\begin{tabular}{ccccc} \toprule
        Ion & Energy & LET at $0^\circ$ & Extra tilt angle & LET$_{\text{eff}}$ \\ 
         & $[$MeV$]$ & $[$MeV/(mg/\cmsq)$]$ & $[^\circ]$ & $[$MeV/(mg/\cmsq)$]$ \\ \midrule
        $^{13}$C$^{4+}$ & 131 & 1.3 & 67 & 3.3 \\
        $^{22}$Ne$^{7+}$ & 238 & 3.3 & 55 & 5.8 \\
        $^{27}$Al$^{8+}$ & 250 & 5.7 & 55 & 9.9 \\
        $^{36}$Ar$^{11+}$ & 353 & 9.9 & 52 & 16.1 \\
        $^{53}$Cr$^{16+}$ & 505 & 16.1 & 38 & 20.4 \\
        $^{103}$Rh$^{31+}$ & 957 & 46.1 & $-$ & $-$ \\
        $^{124}$Xe$^{35+}$ & 995 & 62.5 & $-$ & $-$ \\ \bottomrule
	\end{tabular}
	\caption{Ions, tilt angles, and corresponding Linear Energy Transfers. The effective LET, $\textrm{LET}_\textrm{eff}$, is given by equation~\ref{eqn:LETeff}.}
	\label{tab:LET}
\end{table}

While a full description of the performed tests and results go beyond the scope of this paper, the key result of these measurements showing the energy-dependent cross-sections, based on the same measurement setup as the December proton beam test, are shown below.

\begin{figure}[htbp]
    \centering
    \includegraphics[width=1.\linewidth]{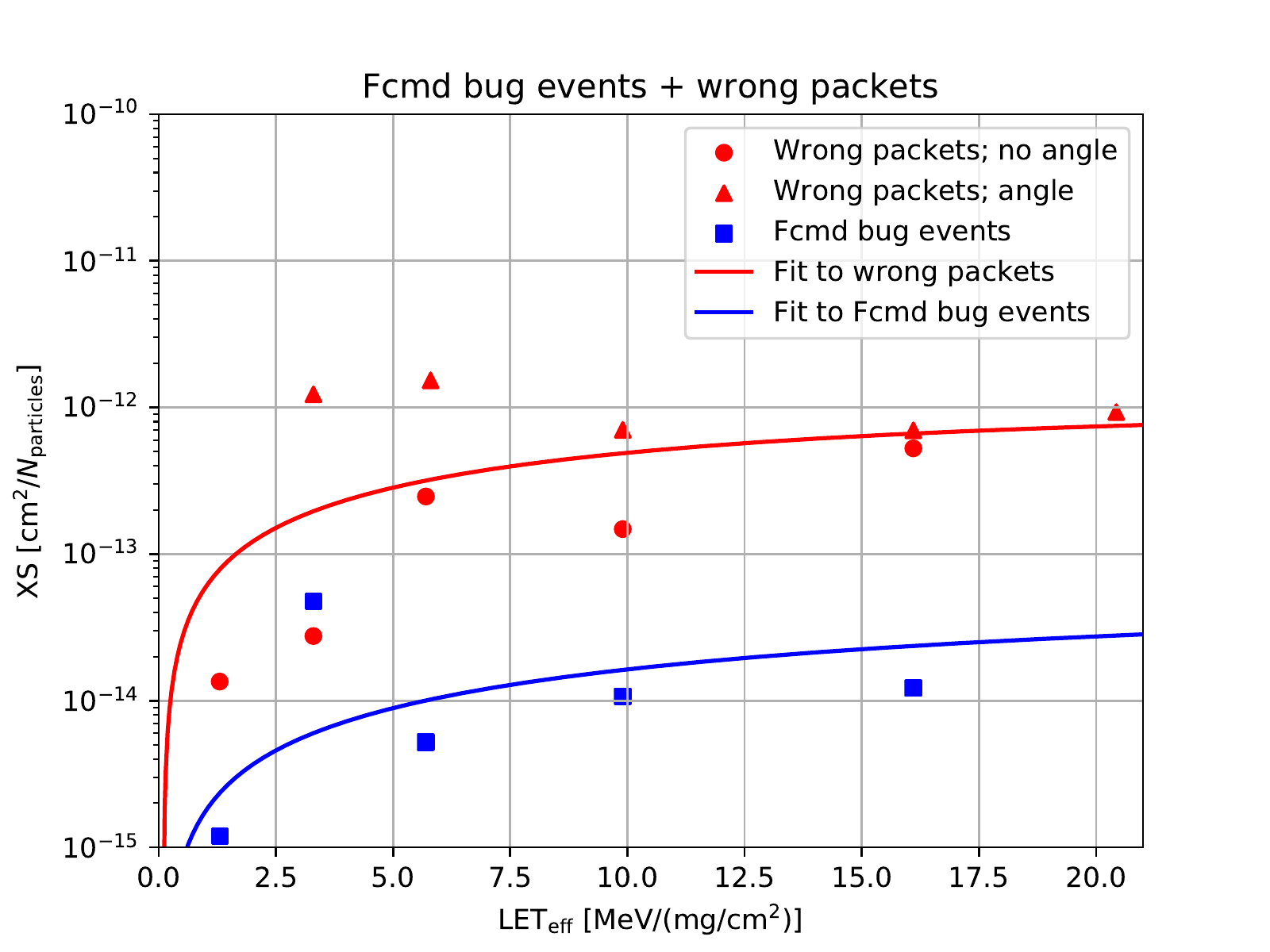}
    \caption{The heavy ion cross-section results from Louvain. The cross-sections for wrong packets (normalised to the total number of packets, with or without an angle) and fast command bugs as well as fits to those cross-sections are shown. The results are plotted as a function of the effective LET, corrected for the angle of the penetrating ions.}
    \label{fig:hi_xs}
\end{figure}

The wrong packet and fast command (``Fcmd'') bug cross-sections as a function of the effective linear energy transfer ($\textrm{LET}_\textrm{eff}$) are shown in figure~\ref{fig:hi_xs}. The ``Fcmd'' bug was identified after chip submission and is due to a small piece of combinatorial logic not being properly triplicated. The sensitive area in the chip of the poorly triplicated logic is very small but it has the nefarious effect of possibly issuing a global chip reset or a BC counter reset if 2~SEUs occur within a few clock cycles of one another. The wrong packet cross-section is given only for physics packets and defines a wrong packet as one with at least one wrong bit in any of the clusters (i.e., the data) -- the packet itself is still required to have a structure consistent with that in table~\ref{tab:packets}. N.B. the effective LET may be calculated from the LET at 0$^\circ$ ($\textrm{LET}_{0^\circ}$), where the ions traverse the ASIC normal to its surface, using:

\begin{equation}
    \textrm{LET}_\textrm{eff} = \frac{\textrm{LET}_{0^\circ}}{\cos\theta} \,,
    \label{eqn:LETeff}
\end{equation}

\noindent where $\theta$ is the angle from the normal.

A fit has been performed to the resulting wrong packet data, both with and without an angle, as well as to the fast command bug cross-sections using a Weibull function:

\begin{equation}
    \sigma(\textrm{LET}_\textrm{eff}) = \sigma_0 \times \left( 1 - \exp\left(\left[\frac{
    \textrm{LET}_\textrm{eff} - L_0}{W}\right]^S\right) \right) \,,
\end{equation}

\noindent where $\sigma_0$ is the cross-section at saturation, $L_0$ is the threshold in effective LET, and $W$ and $S$ are parameters of the Weibull function. For the wrong packet cross-section, the fitted parameters are:

\begin{equation}
    \begin{split}
        \sigma_0 & = 1 \times 10^{-12}~\textrm{cm}^2\textrm{/particle} \\
        L_0      & = 0.1~\textrm{MeV/(mg/cm}^2\textrm{)} \\
        W        & = 22~\textrm{MeV/(mg/cm}^2\textrm{)} \\
        S        & = 1.5
    \end{split} \,,
\end{equation}

\noindent and for the fast command bug cross-section, the fitted parameters are:

\begin{equation}
    \begin{split}
        \sigma_0 & = 5 \times 10^{-14}~\textrm{cm}^2\textrm{/particle} \\
        L_0      & = 0.1~\textrm{MeV/(mg/cm}^2\textrm{)} \\
        W        & = 30~\textrm{MeV/(mg/cm}^2\textrm{)} \\
        S        & = 1.2
    \end{split} \,.
\end{equation}

\noindent These fits are also shown in figure~\ref{fig:hi_xs}.

The resulting Weibull parameters may be used to obtain the cross-sections in the LHC outer tracker environment, which are found to be $4\times10^{-19}$~\cmsq/proton and $2.4\times10^{-20}$~\cmsq/proton for the wrong packet and fast command bug cross-sections, respectively.

As noted in figure~\ref{fig:hi_xs}, the wrong packet cross-section is normalised to the total number of physics packets. Given each ASIC from the proton beam test in December (which had a similar readout routine as Louvain) measured $\mathcal{O}(5\times10^8)/48 \sim \mathcal{O}(10^7)$~packets and only ran for 25\% as long for a single ASIC, this yields a total number of packets\footnote{This approximate number is confirmed from actual data taking: each point for the wrong packet cross-sections presented in figure~\ref{fig:hi_xs} includes a normalisation of $10^6$--$10^7$~packets.} $N_\textrm{packets} \sim \mathcal{O}(10^6)$. Additionally, wrong packets tend to have at most one bit flip, and so $N_\textrm{packets}/n_\textrm{bit-flips} \sim \mathcal{O}(1)$. Multiplying these factors onto the wrong packet cross-section yields a cross-section $\mathcal{O}(10^{-13})$~\cmsq/proton. This number is very consistent with the SEU cross-sections for bit-flips in physics packet clusters measured from the proton beam tests described in section~\ref{sec:res_TRIUMF_phys}, serving as an excellent consistency check.

The fast command bug cross-section is \emph{very} small -- we can estimate the expected number of fast command bugs in a given year. The hadron flux calculation follows the same prescription as equation~\ref{eqn:flux}. With 31,536,000~s/year (the detector will not run the entire year, so this represents a worst-case scenario) and $\sim\!230,000$~ABCStar ASICs in the ITk, the rate of fast command bugs is:

\begin{equation}
    \begin{split}
        \textrm{Rate}_\textrm{Fcmd bugs} & = (230,000)~\textrm{ASICs} \times 31,536,000~\textrm{s/year} \\
        & \,\,\,\,\,\,\,\, \times \mathcal{O}(10^7)~\textrm{hadrons/\cmsq/s} \times (2.4 \times 10^{-20})~\textrm{\cmsq/p/ASIC} \\
        & = \mathcal{O}(2)~\textrm{Fcmd bugs/year} \,.
    \end{split}
\end{equation}

\noindent The rate of fast command bugs is also \emph{very} small. Moreover, even when the fast commands are ASIC resets (as what was measured by the cross-section measurement), reconfiguring a single segment takes \unit[6]{ms}, and so the data loss due to the fast commands bugs may be concluded to be negligible.

\section{Conclusion}

We have characterised the performance ABCStar ASICs Versions 0 and 1 when irradiated with protons at TRIUMF, Canada. This performance includes measurements of the HPR reset and LCB unlock rates (including those reported in the firmware), the physics packet and register read SEU cross-sections, and the digital currents drawn by the chips.

No HPR reset or LCB unlock events were measured for any of the chips included as part of the proton beam tests at TRIUMF. The SEU cross-sections for bit-flips in physics packet clusters were measured to be $(3.69 \pm 0.03) \times 10^{-12}$~\cmsq/p and $(5.32 \pm 0.02) \times 10^{-12}$~\cmsq/p for V0 and V1 chips, respectively. The corresponding error rate is $\mathcal{O}(10^{-10})$~errors/event/ABCStar during normal operation, which is significantly smaller than the noise occupancy due to thermal noise. The SEU cross-section for bit-flips in register read data was measured to be $(1.24 \pm 0.12) \times 10^{-13}$~\cmsq/p for the V0 chip. No SEUs were measured in register read data for the V1 chips, corresponding to a 95\% upper confidence bound of $1.17 \times 10^{-16}$~\cmsq/p on the cross-section -- this result validates the excellent protection conferred by the TMR implemented for V1 registers. Additionally, results obtained with and without the glitch filter enabled were found to be in agreement with each other. As the glitch filter counters glitches on the \unit[40]{MHz} clock input pad and because this input pad was an ideal source for both SEU beam test campaigns, these results are expected.

Measurements of the digital current drawn by the ABCStar ASICs has confirmed the validity of pre-irradiating the chips to guard against high current loads during runtime. This behaviour is consistent with what was observed for the prototype ABC130 ASICs, demonstrating it to be a stable feature of the technology used to manufacture both the ABC130 and ABCStar ASICs.

Complementary measurements have been performed for ABCStar ASICs irradiated with heavy ions in the cyclotron facility at Louvain-la-Neuve, Belgium. The cross-section for physics packets with wrong clusters obtained from heavy ion measurements were found to agree with the magnitude of the cross-section for bit-flips in physics packet clusters obtained from proton beam measurements. Additionally, the cross-section for fast command bugs obtained from heavy ion measurements was shown to have a negligible impact on detector operation and data taking.

\acknowledgments
\addcontentsline{toc}{section}{Acknowledgements}

This work was supported by the Canada Foundation for Innovation and the Natural Sciences and Engineering Research Council of Canada as well as the Alexander von Humboldt Foundation. We acknowledge the financial support by the Federal Ministry of Education and Research of Germany.

\Urlmuskip=0mu plus 1mu\relax
\bibliographystyle{JHEP}
\bibliography{bibliography.bib}

\appendix
\section{Dose and fluence measurements from proton tests}
\label{app:res_TRIUMF_dose_meas}

\begin{figure}[htbp]
    \centering
    \begin{subfigure}{.49\textwidth}
        \centering
        \includegraphics[width=1.\linewidth]{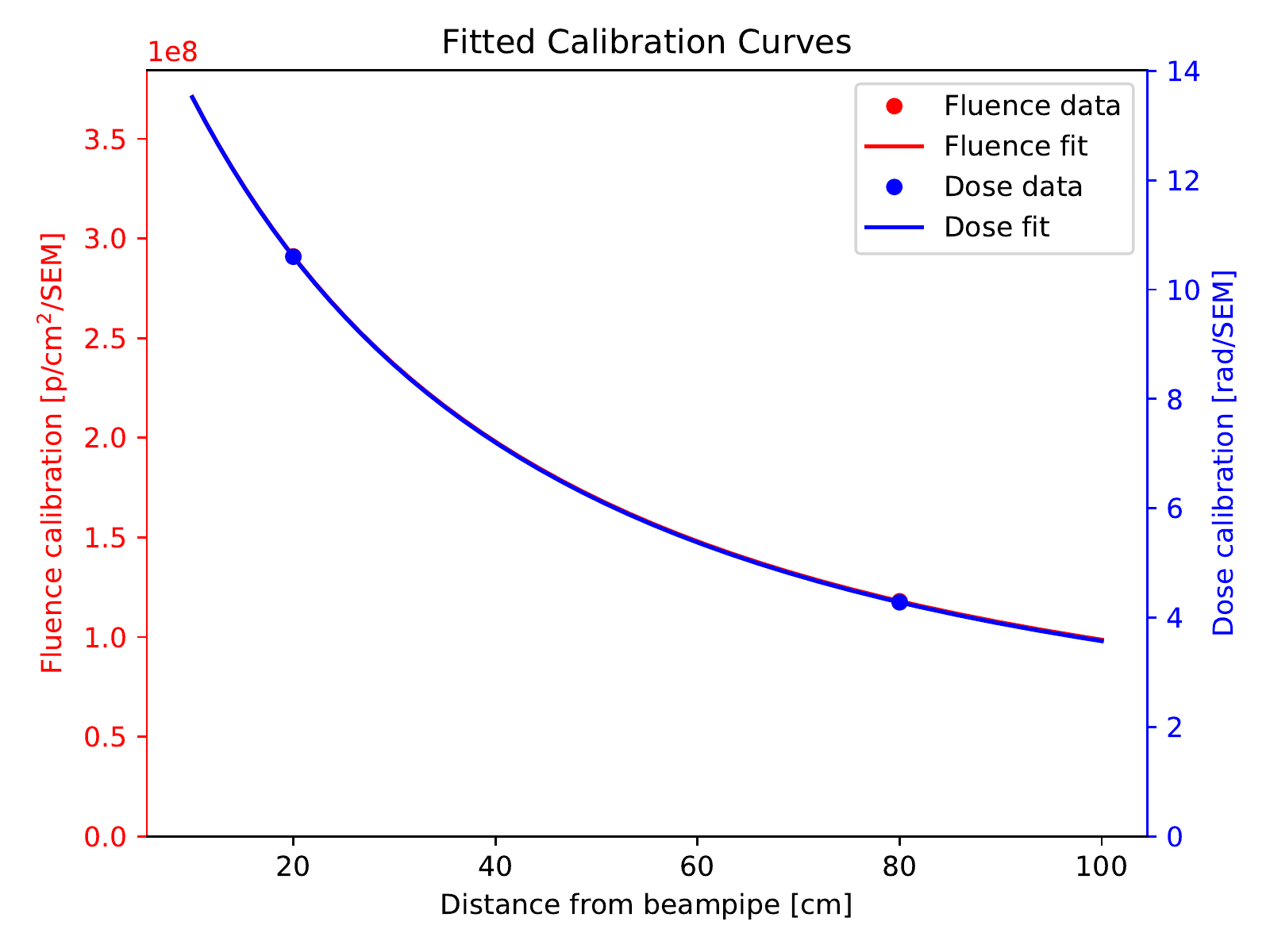}
        \caption{August 2020 calibrations}
        \label{fig:proton_calib_aug}
    \end{subfigure}
    \hfill
    \centering
    \begin{subfigure}{.49\textwidth}
        \centering
        \includegraphics[width=1.\linewidth]{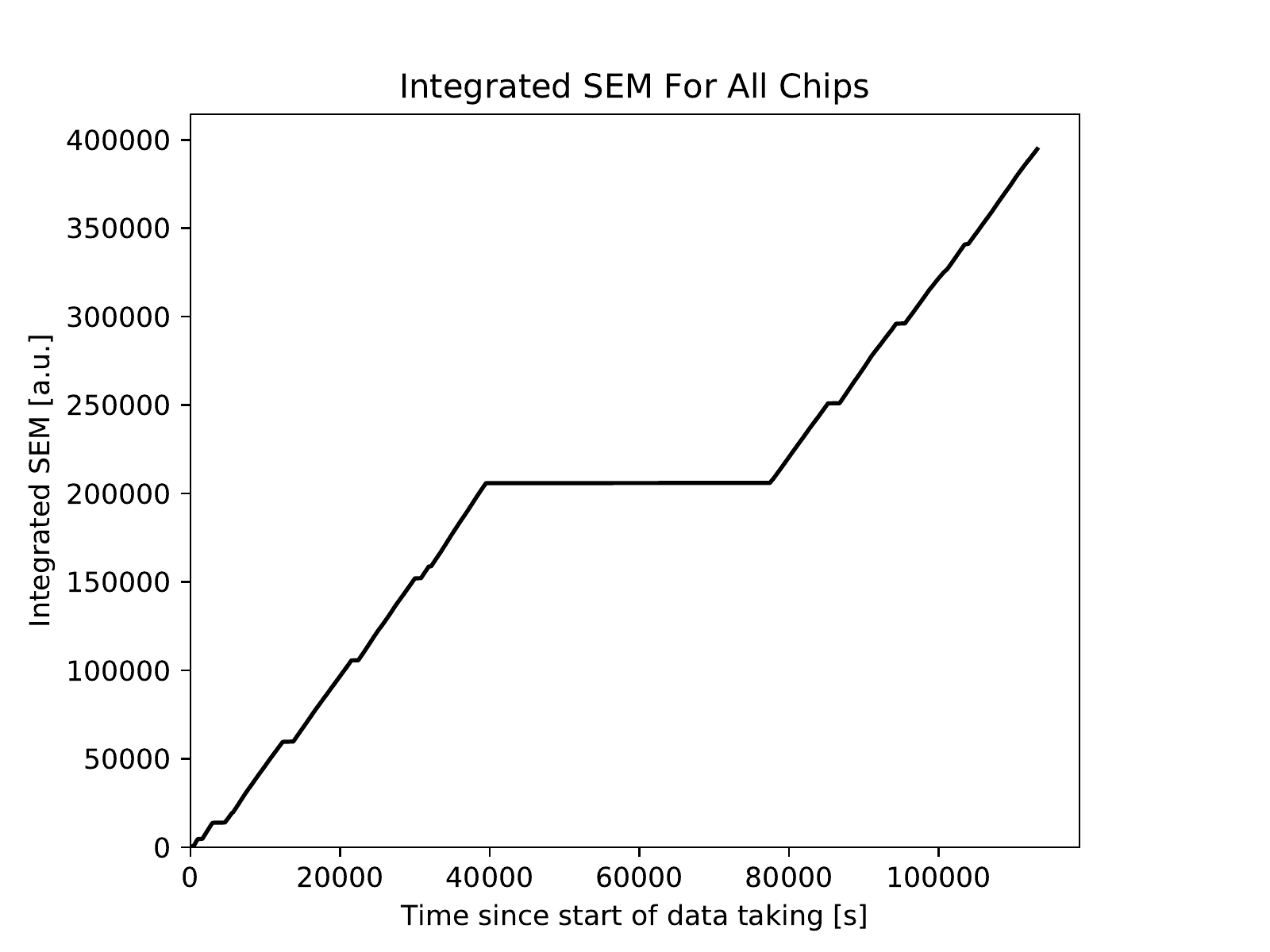}
        \caption{August 2020 SEM counts}
        \label{fig:proton_sem_aug}
    \end{subfigure} \\
    \centering
    \begin{subfigure}{.49\textwidth}
        \centering
        \includegraphics[width=1.\linewidth]{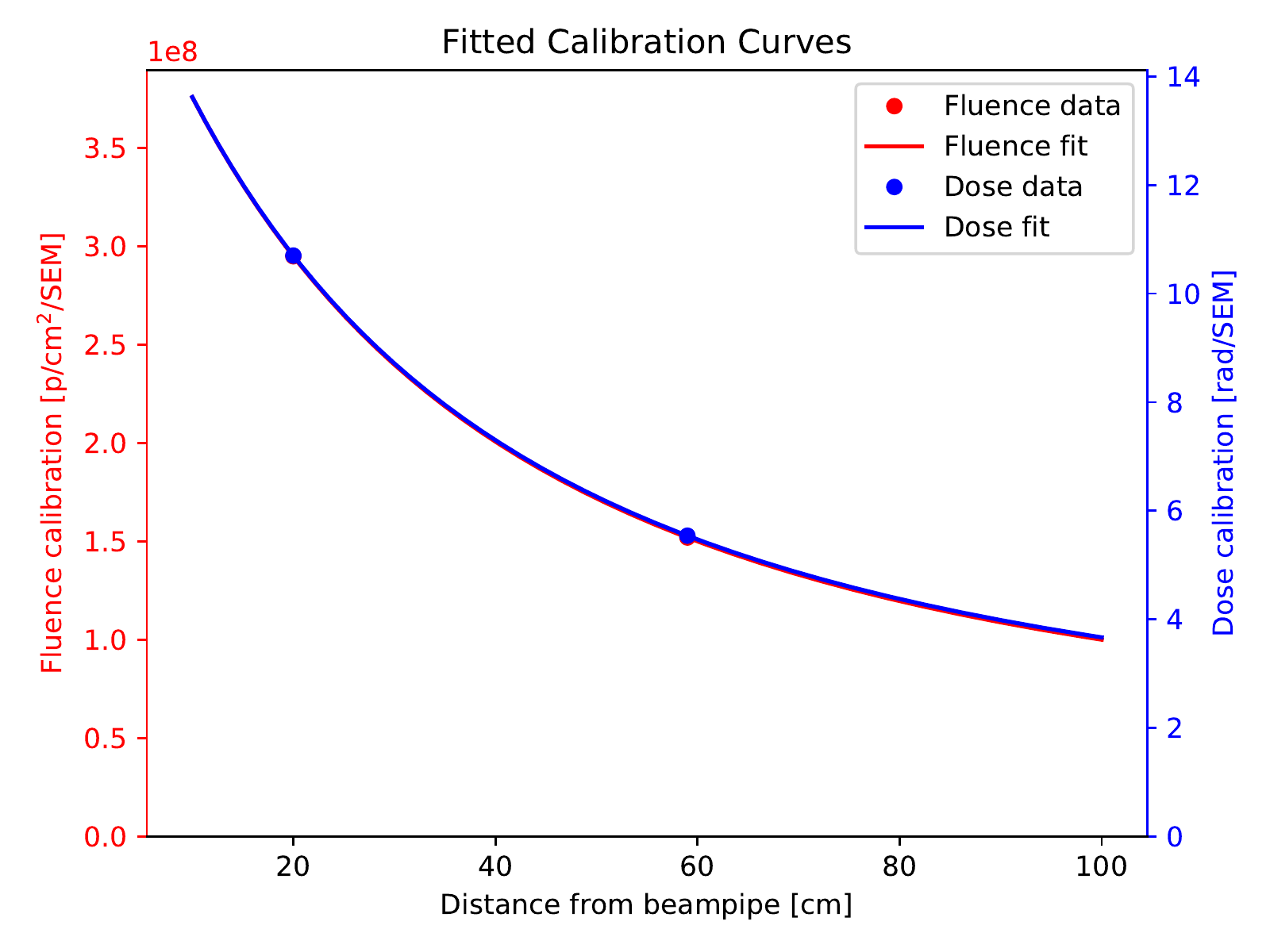}
        \caption{December 2020 calibrations}
        \label{fig:proton_calib_dec}
    \end{subfigure}
    \hfill
    \centering
    \begin{subfigure}{.49\textwidth}
        \centering
        \includegraphics[width=1.\linewidth]{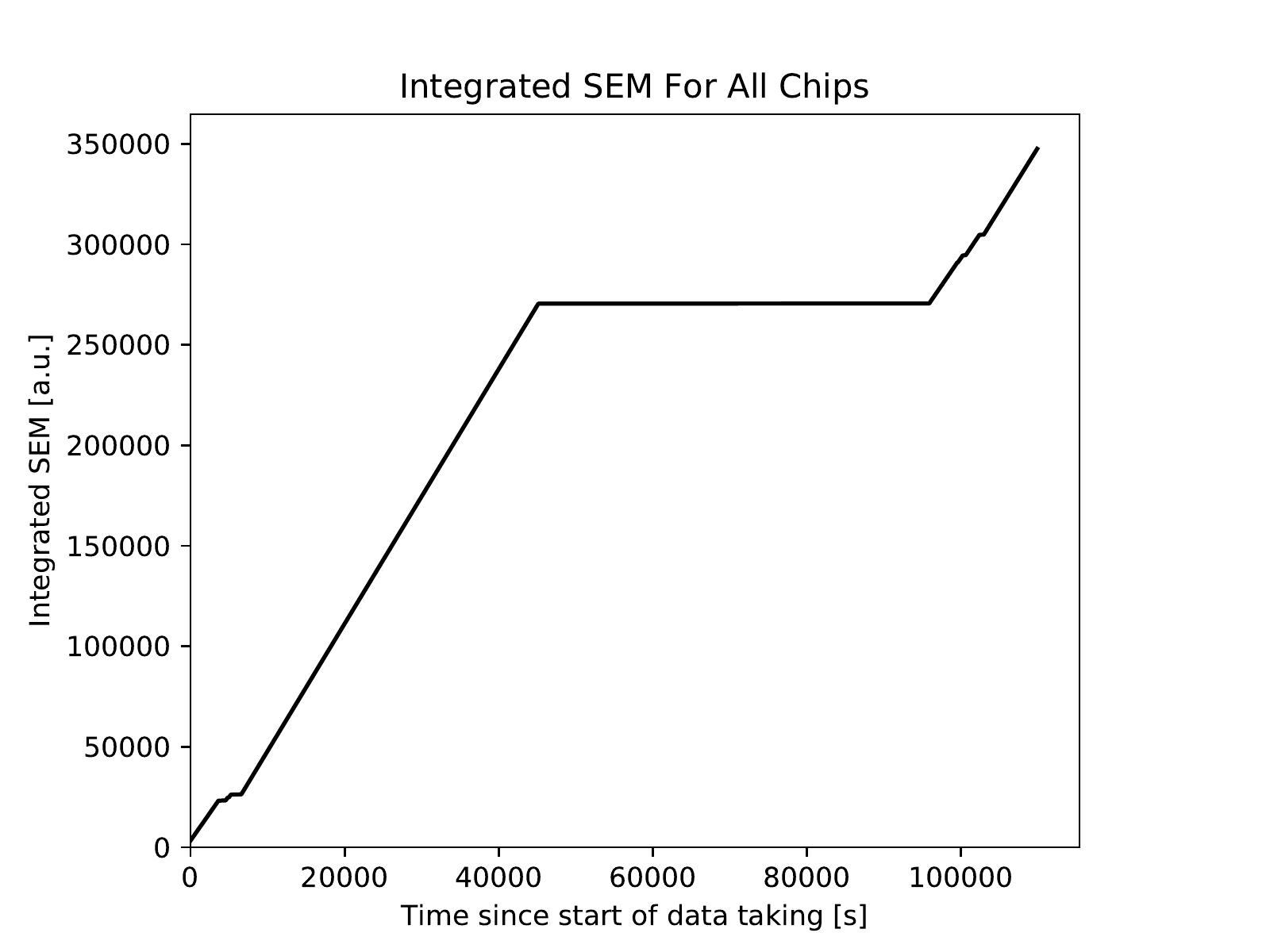}
        \caption{December 2020 SEM counts}
        \label{fig:proton_sem_dec}
    \end{subfigure} \\
    \caption{The results of the TRIUMF PIF irradiations. (a) and (c) show the fluence and dose delivered per SEM as a function of radial distance from the beampipe for the August and December 2020 proton beam tests, respectively. The red data points and curves lie underneath the blue data points and curves. (b) and (d) show the integrated SEM for all chips as a function of time since the start of data taking for the August and December 2020 proton beam tests, respectively.}
    \label{fig:proton_rad}
\end{figure}

Using the calibrations provided by the TRIUMF PIF, the fluence and dose delivered per Secondary Emission Monitor (SEM) counts as a function of the radial distance from the beampipe was fit. The calibration data and resulting fits as well as the integrated SEM as a function of time is shown in figure~\ref{fig:proton_rad}. The total integrated dose or fluence at a time $t$ since the start of data taking, $\phi(t)$, is calculated as:

\begin{equation}
    \phi(t) = \textrm{Calibration}(R) \times \textrm{SEM}(t) \,,
\end{equation}

\noindent where $\textrm{Calibration}(R)$ is the value of the calibration curve for a chip at radial distance $R$ from the beampipe and $\textrm{SEM}(t)$ is the integrated SEM.

\FloatBarrier

\end{document}